\documentclass[pdflatex,sn-nature]{sn-jnl}

\usepackage{graphicx}
\graphicspath{{./}{figs/}}
\usepackage{amsmath,amssymb,bm}
\usepackage{booktabs}
\usepackage{siunitx}
\usepackage{multirow}
\usepackage{array}
\usepackage{xcolor}
\usepackage{url}
\usepackage{caption}
\usepackage{tabularx}
\usepackage{float}
\usepackage{import}
\usepackage{soul}

\title[XQNN for multi-material topology optimization]{Explainable quantum neural networks for multi-material topology optimization}

\author[1]{\fnm{Dahyun} \sur{Joo}}
\author[2]{\fnm{Naruethep} \sur{Sukulthanasorn}}
\author[2,3]{\fnm{Kenjiro} \sur{Terada}}
\author*[1,4,5]{\fnm{Do-Nyun} \sur{Kim}}\email{dnkim@snu.ac.kr}

\affil[1]{\orgdiv{Department of Mechanical Engineering}, \orgname{Seoul National University}, \orgaddress{\city{Seoul}, \postcode{08826}, \country{Republic of Korea}}}
\affil[2]{\orgdiv{International Research Institute of Disaster Science}, \orgname{Tohoku University}, \orgaddress{\city{Sendai}, \postcode{980-8572}, \country{Japan}}}
\affil[3]{\orgdiv{Department of Civil and Environmental Engineering}, \orgname{Tohoku University}, \orgaddress{\city{Sendai}, \postcode{980-8579}, \country{Japan}}}
\affil[4]{\orgdiv{Institute of Advanced Machines and Design}, \orgname{Seoul National University}, \orgaddress{\city{Seoul}, \postcode{08826}, \country{Republic of Korea}}}
\affil[5]{\orgdiv{Institute of Engineering Research}, \orgname{Seoul National University}, \orgaddress{\city{Seoul}, \postcode{08826}, \country{Republic of Korea}}}

\begin{document}

\abstract{We propose an explainable quantum neural network for multi-material topology optimization, XQNN, that determines both load-carrying structural layout and material type assignment for given boundary/loading conditions. Intermediate solution histories are first converted into element-wise strain energy, sensitivity, density, and Sobel boundary descriptors. Then, they are encoded in a ten-qubit circuit and qubit-wise $Z$ observables are mapped onto material type labels. Trained only on two-dimensional topology optimization histories obtained with a fixed mesh resolution, XQNN can be generalized to handle out-of-distribution boundary/loading conditions, progressively refined high-resolution meshes, and voxel-wise three-dimensional problems without additional training. We find that it is important to preserve qubit-wise observables and add boundary information for improving the optimization accuracy, and certain observables have consistent links to load paths, material type regions, and interfaces, demonstrating their usability as auditable mechanics-facing variables.}

\maketitle

\section{Introduction}

Topology optimization (TO) distributes material within a prescribed design domain so that a structural objective, most commonly compliance, is optimized under equilibrium and resource constraints. It has matured through homogenization, density-based formulations, level-set methods, deformable mesh strategies, and broad computational implementations \cite{BendsoeKikuchi1988,BendsoeSigmund2003,Sigmund2001,SigmundMaute2013,DeatonGrandhi2014,VanDijkEtAl2013,AllaireJouveToader2004,JungKim2025DeformableMesh}. Multi-material topology optimization (MMTO) considers the assignment of material types in addition to the usual layout decision. This coupled layout--type decision creates material--material interfaces, void--material interfaces, and a larger combinatorial design space, motivating the development of specialized formulations for multi-material structural design \cite{TavakoliMohseni2014,ZuoSaitou2017,VermaakEtAl2014,WangEtAl2015MMLevelSet,MirzendehdelSuresh2015,LiKim2018,ZhengEtAl2024MMTOCode}.

Learning-assisted TO is attractive in this setting because repeated MMTO calculations are computationally demanding when boundary conditions, loading scenarios, mesh resolutions, or design requirements change. Convolutional neural networks, generative models, multi-stage predictors, reinforcement-learning strategies, and super-resolution based accelerators have demonstrated that near-optimized topologies can be inferred directly from physical fields, early optimizer states, or problem specifications \cite{WoldsethEtAl2022,YuEtAl2019,NieEtAl2021,BieleckiEtAl2021,XueEtAl2021CNN,BrownEtAl2022,GaoEtAl2024,LimEtAl2024SuperResolution}. While such learned predictors can substantially accelerate the design process, they also impose an interpretability requirement: producing a plausible material map is insufficient if the model cannot account for the signals that support its prediction. A mechanics surrogate should make clear whether its decision is driven by load-carrying regions, material interfaces, or features that separate material types, and these signals should remain accessible when the test case departs from the training distribution. This requirement aligns with a broader view of interpretable and physics-informed machine learning, where physical consistency is regarded as a modeling criterion \cite{Rudin2019,RoscherEtAl2020,ArrietaEtAl2020XAI,KarniadakisEtAl2021}. Although generic post-hoc tools such as SHAP and gradient localization are useful in many domains \cite{LundbergLee2017SHAP,SelvarajuEtAl2017GradCAM}, their explanatory vocabulary in compact numerical surrogates is often inherited from the underlying architecture.

Parameterized quantum circuits (PQCs), often referred to as quantum neural networks (QNNs) in learning applications, provide a natural route to such explicitly accessible variables through measured expectation values of selected observables. A PQC encodes classical inputs into a quantum state, transforms the state through trainable unitary operations, and reads out the resulting state by bounded expectation values associated with selected qubits and measurement operators. In this work, each measured expectation value is termed a quantum observable channel; specifically, each channel corresponds to a qubit-wise Pauli-Z expectation, $\langle Z_i\rangle$, obtained after feature encoding and trainable circuit transformation. 

This observable-channel formulation is well suited to MMTO because the assignment of material types is not governed solely by scalar density values. It depends on competing material types, interface adjacent elements, and coupled mechanical descriptors. Compressing all measured qubit responses into a single scalar can obscure circuit-level information generated through superposition and entangling operations. Retaining qubit-wise observables instead preserves multiple bounded response channels, giving the readout layer independent degrees of freedom for multi-material classification while keeping the intermediate variables explicitly measurable.

This perspective extends earlier work on quantum-assisted topology optimization. PQCs have been shown to accelerate topology prediction by using optimizer-derived strain energy, sensitivity, and density descriptors from early TO iterations \cite{SukulthanasornTerada2026}. The binary-topology formulation used fixed early optimizer snapshots and converted a scalar quantum response into topology decisions through an empirical threshold. Here, we propose an explainable quantum neural network (XQNN) that retains the accelerator-oriented motivation of that approach but adapts it to MMTO. More specifically, fixed early snapshots are replaced by mechanically defined milestones, $k_{\rm em}$ and $k_{\rm em}+5$, to provide a richer post-emergence history, together with the prediction decision milestone $k_{\rm pd}$; a scalar quantum response is replaced by qubit-wise $Z$ observables followed by a softmax readout; and a Sobel boundary hint is introduced to provide interface sensitive information. Variational quantum latent encoding has also been explored for TO, where quantum circuits form compact latent representations of design information for optimization-oriented prediction \cite{Tabarraei2025VQLETO}. These studies motivate the central question of this work: can a compact PQC generate measured observables that remain predictive after readout and can it also be interrogated in mechanics-facing terms?

\section{Results}

The learning task is element-wise material classification for MMTO layouts generated by the method of moving asymptotes (MMA) where each finite element is assigned to one of the available solid material types or to void. We denote the number of solid material types by $M$ and study two-, three-, and four-material cases. Separately, the grayscale ratio $G(k)$ is used only to quantify the discreteness of an evolving material field during MMA optimization and to select intermediate iterations, where $k$ denotes the MMA iteration index. The problems and dataset are organized into non-overlapping roles (Tables~S1 and S2). Training problems $T_1$--$T_3$ are used to fit the circuit and readout parameters, validation problems $V_1$--$V_3$ are used for model selection and early stopping, and out-of-distribution (OOD) test problems $O_1$--$O_4$ are used to evaluate the performance of the selected model under boundary/loading conditions unseen during training. The high-resolution transfer cases $\mathit{HR}_1$--$\mathit{HR}_3$, defined by progressively refined meshes, and the three-dimensional transfer cases $\mathit{3D}_1$--$\mathit{3D}_4$ are designed to demonstrate the generalization performance of XQNN. Across all material layout plots, colors represent the material type: blue, orange, yellow, and green for material types 1, 2, 3, and 4, respectively, with white for void.

\subsection{Model architecture of the XQNN}

\begin{figure}[p]
\centering
\includegraphics[width=\linewidth, trim={0.0cm 6.0cm 0.0cm 0.0cm}, clip]{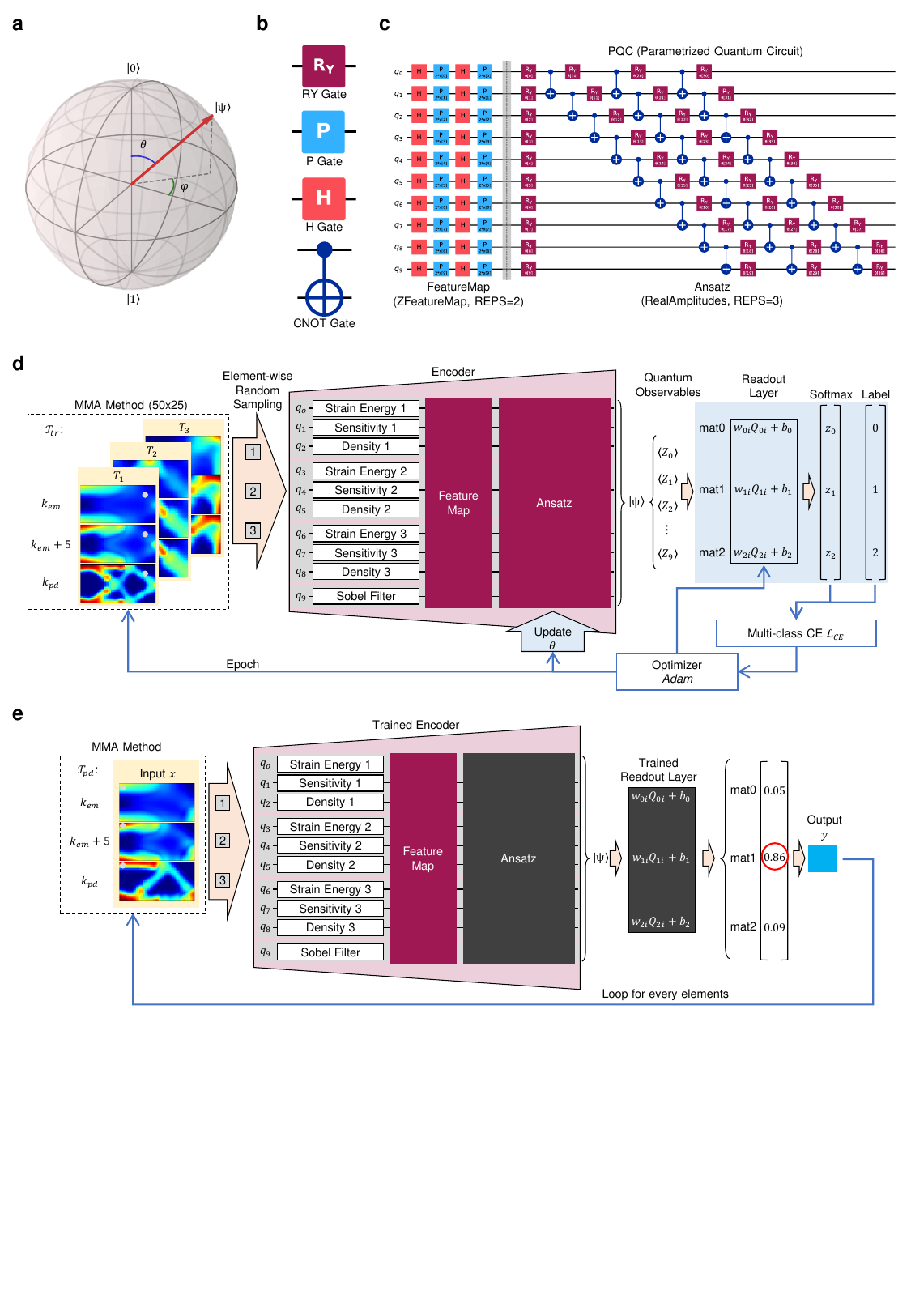}
\caption{\textbf{XQNN workflow for multi-material topology prediction.} \textbf{a}, Bloch-sphere representation of a single qubit. \textbf{b}, Gate primitives used by the implemented circuit: Hadamard and phase gates in the ZFeatureMap, and trainable $R_Y$ rotations with CNOT entanglers in the RealAmplitudes ansatz. \textbf{c}, Ten-qubit circuit structure. The first nine qubits encode strain energy, sensitivity, and density descriptors from $k_{\rm em}$, $k_{\rm em}+5$ and $k_{\rm pd}$, while the tenth qubit encodes a Sobel boundary hint. The resulting state is measured through qubit-wise $Z$ observables. \textbf{d}, Training path from selected MMA iterations to class balanced element-wise sampling and multi-class cross-entropy optimization. \textbf{e}, Prediction path in which a new MMA run is stopped at $k_{\rm pd}$, each element is encoded by the trained circuit, and the trained readout layer returns material type probabilities.}
\label{fig:workflow}
\end{figure}

The XQNN representation keeps the quantum output as a dictionary of measured observables rather than collapsing it into one scalar. A pure qubit can be written as
\begin{equation}
|\psi(\vartheta,\varphi)\rangle=
\cos\frac{\vartheta}{2}|0\rangle+
e^{i\varphi}\sin\frac{\vartheta}{2}|1\rangle,
\label{eq:bloch_results}
\end{equation}
where $\vartheta$ and $\varphi$ are the polar and azimuthal angles on the Bloch sphere (Figure~\ref{fig:workflow}a). After feature encoding and ansatz transformation, each qubit is measured through a $Z$ expectation,
\begin{equation}
\langle Z_i\rangle=p_i(0)-p_i(1),
\label{eq:z_expectation_results}
\end{equation}
which reports the population contrast of qubit $i$. This bounded expectation value is the quantum observable channel used by the readout layer.

Hadamard and phase gates form the ZFeatureMap, and trainable $R_Y$ rotations with CNOT entanglers form the RealAmplitudes ansatz (Figure~\ref{fig:workflow}b). The Hadamard and phase gates introduce descriptor-dependent quantum phases, whereas the $R_Y$ rotations and CNOT entanglers mix the encoded state before the final $Z$ measurements. The MMTO histories and preprocessing used to construct the descriptors are described first in Supplementary Notes~1 and 2, and the corresponding gate-level matrices and circuit ordering are given afterward in Supplementary Note~3.

Figure~\ref{fig:workflow}c defines the ten-qubit XQNN input. Each element is represented by ten descriptors: strain energy ($E_e$), sensitivity ($S_e$), and density ($D_e$) sampled at the emergence iteration $k_{\rm em}$, five MMA steps later at $k_{\rm em}+5$, and the prediction decision iteration $k_{\rm pd}$, together with a Sobel boundary hint ($B_e$) computed from the prediction stage density field. The descriptor vector is then
\begin{equation}
\begin{aligned}
\mathbf{x}_e =
\big[&
E_e(k_{\rm em}),\, S_e(k_{\rm em}),\, D_e(k_{\rm em}),\\
&
E_e(k_{\rm em}+5),\, S_e(k_{\rm em}+5),\, D_e(k_{\rm em}+5),\\
&
E_e(k_{\rm pd}),\, S_e(k_{\rm pd}),\, D_e(k_{\rm pd}),\, B_e(k_{\rm pd})
\big]^T .
\end{aligned}
\label{eq:descriptor_results}
\end{equation}
Adapting the optimizer history strategy \cite{SukulthanasornTerada2026} to MMTO, the same relative MMA milestones are used in both training and prediction. Here, $k_{\rm em}$ replaces a fixed first early snapshot as the first mechanically meaningful emergence state, $k_{\rm em}+5$ replaces the next early snapshot as a richer post-emergence state, and $k_{\rm pd}$ provides the early stopped prediction decision state. During inference, a new MMA run is stopped at $k_{\rm pd}$, and descriptors from $k_{\rm em}$, $k_{\rm em}+5$, and $k_{\rm pd}$ are passed to the trained XQNN.

The raw descriptor values are normalized with a training set min--max transform before quantum encoding. Ground-truth (GT) material type labels are refined with a $3\times3$ mode filter, and the scaling transform is fitted independently for each material type setting because the physical ranges of strain energy, sensitivity, density, and Sobel boundary hint values differ across the two-, three-, and four-material simulations. For material setting $M$ and descriptor/qubit index $q$,
\begin{equation}
\tilde{x}_{e,q}=\operatorname{clip}\!\left(\mathcal{S}_{M,q}(x_{e,q}),0,1\right),
\qquad
\alpha_{e,q}=\pi\tilde{x}_{e,q}.
\label{eq:angle_clip_results}
\end{equation}
The pre-fitted transform is reused without refitting for OOD, high-resolution, and three-dimensional inference, and clipping keeps the encoded angle in the valid feature map range $[0,\pi]$.

The normalized descriptor is encoded by the ZFeatureMap with REPS=2, transformed by the RealAmplitudes ansatz with REPS=3, and measured as
\begin{equation}
\mathbf{Q}_e=\big(\langle Z_0\rangle_e,\langle Z_1\rangle_e,\ldots,\langle Z_9\rangle_e\big)^T.
\label{eq:observable_results}
\end{equation}
During training (Figure~\ref{fig:workflow}d), class balanced element samples from the selected MMA histories optimize the circuit parameters and readout weights. During prediction (Figure~\ref{fig:workflow}e), a new topology optimization run is stopped at $k_{\rm pd}$, each element is encoded by the trained circuit, and the readout layer maps $\mathbf{Q}_e$ to material type probabilities,
\begin{equation}
\mathbf{z}_e=\mathbf{W}^T\mathbf{Q}_e+\mathbf{b},
\qquad
p_{e,c}=\frac{\exp(z_{e,c})}{\sum_{r=0}^{C-1}\exp(z_{e,r})},
\qquad
\hat{y}_e=\arg\max_c p_{e,c}.
\label{eq:softmax_results}
\end{equation}
The readout has $C=M+1$ classes for $M$ solid material types plus void. The remaining implementation details, including the exact statevector evaluation, training schedule, and parameter counts, are given in Methods and Supplementary Note~3.

\subsection{Model training and validation}

\begin{figure}[p]
\centering
\includegraphics[width=\linewidth, trim={0.0cm 3.5cm 0.0cm 0.0cm}, clip]{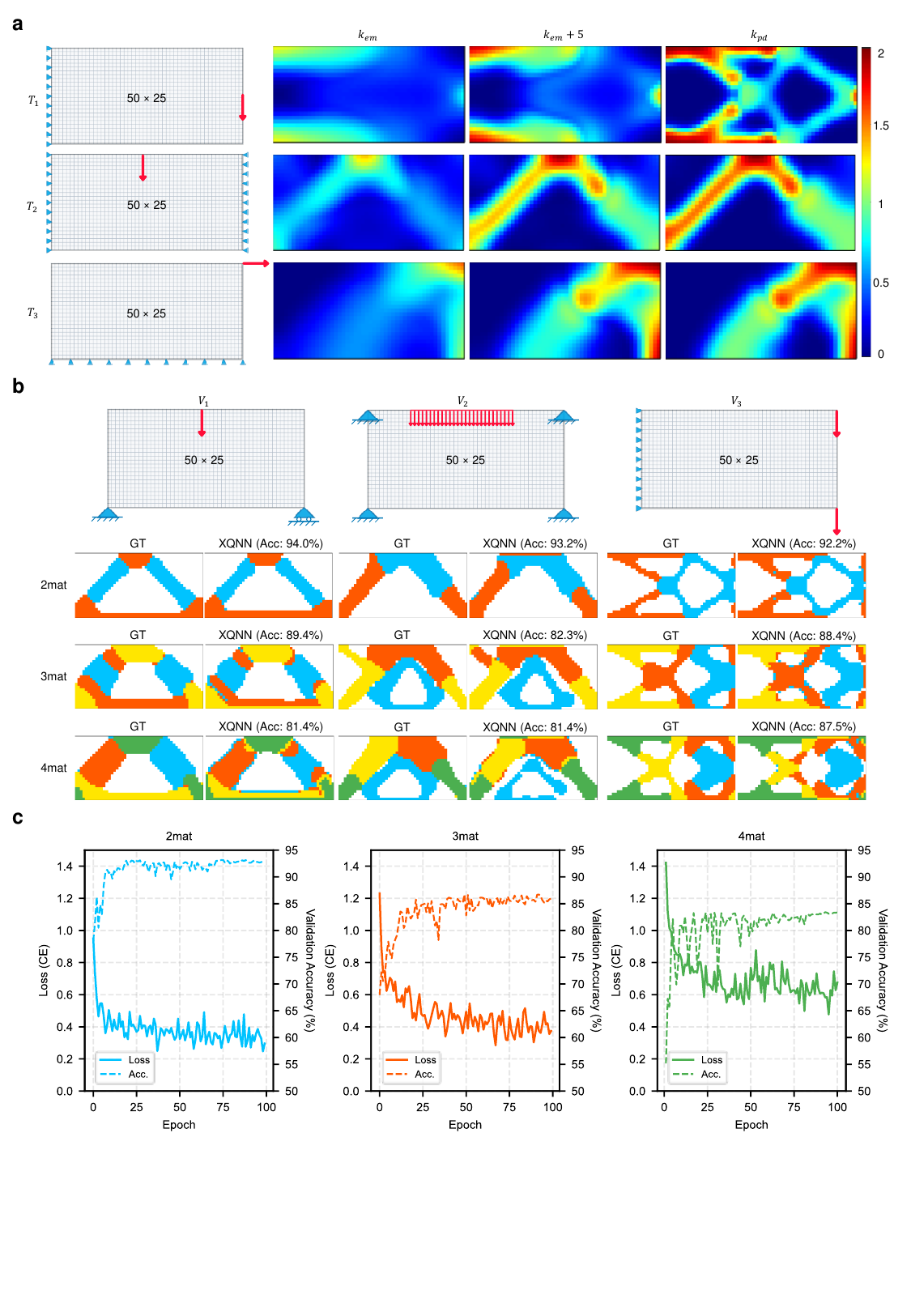}
\caption{\textbf{Training, validation, and convergence behavior.} \textbf{a}, Training boundary/loading conditions $T_1$--$T_3$ and selected MMA fields at $k_{\rm em}$, $k_{\rm em}+5$, and $k_{\rm pd}$ used as XQNN input descriptors. \textbf{b}, Validation conditions $V_1$--$V_3$ with GT and XQNN-inferred material type layouts; validation is used only for model selection and early stopping. \textbf{c}, Training loss and validation accuracy plotted with common axes; dashed lines denote the mean validation accuracy over $V_1$--$V_3$.}
\label{fig:training}
\end{figure}

We train the model using MMA histories of two-dimensional MMTO problems solved with a $50\times25$ mesh. Three boundary/loading conditions are used to obtain the training dataset: ($T_1$) under a vertical point load at the right edge while fixed at the left edge, ($T_2$) under a vertical point load at the center of the top edge while fixed at both left and right edges, and ($T_3$) under a horizontal point load at the upper right corner while fixed at the bottom edge (Figure~\ref{fig:training}a). The optimizer fields sampled at $k_{\rm em}$, $k_{\rm em}+5$, and $k_{\rm pd}$ are used for training after being converted into element-wise XQNN descriptors.

The validation set is constructed using three additional boundary/loading conditions: ($V_1$) under a vertical point load at the center of the top edge while simply supported at the bottom corners, ($V_2$) under a distributed vertical load on the central part of the top edge while fixed at the four corners, and ($V_3$) under two vertical point loads at the right corners while fixed at the left edge (Figure~\ref{fig:training}b). Validation is used only for model selection and early stopping. The validation predictions retain the dominant load paths and material type regions with high material type accuracy across $V_1$--$V_3$, showing that the selected intermediate fields carry useful information for boundary/loading conditions that are not used for fitting the hyperparameters. The two- and three-material predictions reproduce the main structural members well, with some local shifts near the void--solid or material--material boundaries. The four-material case is visually less clean, but the main load path and the largely clustered material regions remain recognizable. The accuracy ranges from 81.4\% to 94.0\%.

The model can be trained stably regardless of the number of material types used (Figure~\ref{fig:training}c). The training loss is evaluated on the class balanced random subset sampled at each epoch, while the validation accuracy is calculated by taking the mean of material type overlap for all elements in $V_1$--$V_3$. Nevertheless, convergence of the loss function becomes slower and the validation accuracy decreases as the number of material types increases, because it gets harder for the same compact observable dictionary to separate more material types and material--material interfaces.

Increasing the number of training designs further improves or stabilizes the validation behavior, but three training designs already provide sufficient validation accuracy (Figure~S1 and Table~S3). Training loss decreases as more optimization histories are supplied while the mean validation accuracy changes modestly only and can even appear highest when only $k_{\rm pd}$ is used (Figure~S2 and Table~S4). More importantly, high-resolution transfer tests with altered stiffness settings in a two-material case show that using multiple milestones, $k_{\rm em}$, $k_{\rm em}+5$, and $k_{\rm pd}$, produces more stable holes, material paths, and connector regions than using a single late snapshot. Varying the depth of the RealAmplitudes indicates that REPS=3 balances expressivity and stable prediction, while deeper circuits do not consistently improve the validation cases (Figure~S3 and Table~S5). The target grayscale threshold $G_{\rm target}$ controls a trade-off between the computing time and solution accuracy (Figure~S4 and Table~S6). Lowering $G_{\rm target}$ moves $k_{\rm pd}$ closer to the final optimized layout and usually improves validation accuracy, but also reduces the benefit of early stopping. We select $G_{\rm target}=0.20$, 0.15, and 0.10 for the two-, three-, and four-material cases, respectively, as a compromise that retains a clear early stopping benefit while providing sufficient boundary clarity for prediction. Feature space and prediction distribution diagnostics further show that confidence broadens and material type frequency imbalance becomes more pronounced in the four-material case, consistent with its more distributed errors (Figure~S5).

\subsection{Generalization performance}

\begin{figure}[p]
\centering
\includegraphics[width=\linewidth, trim={0.0cm 4.0cm 0.0cm 0.0cm}, clip]{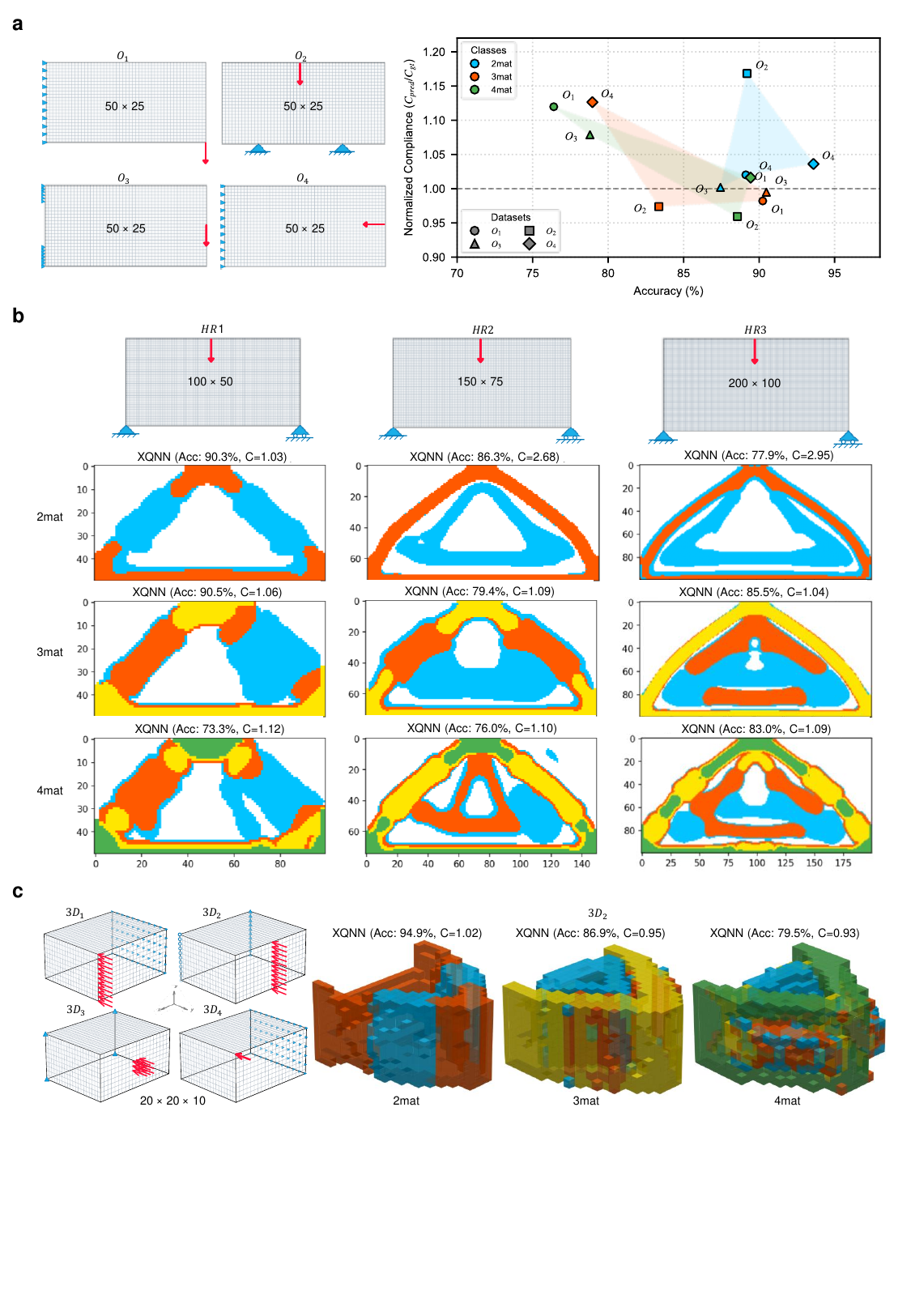}
\caption{\textbf{Generalization performance under OOD, high-resolution, and three-dimensional transfer.} \textbf{a}, OOD tests $O_1$--$O_4$ on the same $50\times25$ mesh and the corresponding scatter plot of material type accuracy versus normalized compliance $C_{\rm pred}/C_{\rm GT}$ for all material settings. \textbf{b}, High-resolution transfer cases $\mathit{HR}_1$($100\times50$), $\mathit{HR}_2$($150\times75$), and $\mathit{HR}_3$($200\times100$), with representative XQNN predictions for two-, three-, and four-material cases. \textbf{c}, Three-dimensional transfer conditions $\mathit{3D}_1$--$\mathit{3D}_4$ and representative $\mathit{3D}_2$ predictions on a $20\times20\times10$ voxel mesh.}
\label{fig:prediction}
\end{figure}

We evaluate the generalization performance of the trained model using two-dimensional OOD boundary/loading conditions, various mesh resolutions, and three-dimensional TO problems (Figure~\ref{fig:prediction}). In addition to pixel-wise accuracy of assigned material types, we calculate the normalized compliance, $C_{\rm pred}/C_{\rm GT}$, indicating the relative stiffness of the inferred structure compared to that of the GT reference solution. The full GT--prediction layout pairs for these problems are provided in Figures~S6--S8 for the two-, three-, and four-material settings, respectively.

\textbf{OOD boundary/loading conditions} We apply four additional boundary/loading conditions, denoted $O_1$--$O_4$, to the same $50\times25$ mesh to assess OOD performance (Figure~\ref{fig:prediction}a). The scatter plot compares the accuracy of material types and normalized compliance for different numbers of materials across four cases. Overall, the two-material case exhibits relatively high accuracy and compliance values close to unity, indicating stable predictive performance. In contrast, the three- and four-material cases show larger problem-dependent variations, as reflected by the broader shaded regions. The case $O_2$ in the two-material TO shows a noticeable deviation in normalized compliance while achieving high accuracy. This turns out to be due to a local shift at the material interface rather than a deficient structural layout (Figure~S6). The other three cases with relatively high values of normalized compliance reveal a similar aspect in that the principal load-carrying skeleton is well predicted by XQNN but local errors in material type assignments particularly near the material interfaces deteriorate the structural stiffness. As the number of material types increases, the agreement between predicted and GT compliance becomes more sensitive to the problem because the number of material interfaces increases naturally.

\textbf{Mesh resolution} We test the generalization performance of the XQNN model by applying it, without additional training, to the $V_1$ case solved using three higher-resolution meshes, $\mathit{HR}_1$($100\times50$), $\mathit{HR}_2$($150\times75$), and $\mathit{HR}_3$($200\times100$) (Figure~\ref{fig:prediction}b). Note that this test evaluates the mesh-scale transfer of our element-wise observable model, rather than the fixed-domain mesh convergence. Across resolutions, the inferred structures retain arch-like load paths and central voids, indicating that XQNN is not dependent on the mesh resolution at which it is trained. For the three- and four-material cases, XQNN provides stable solutions with good material type accuracies (ranging from 73\% to 91\%) and normalized compliance values close to one. No clear mesh dependency is observed. However, the two-material cases become less stable, particularly when higher mesh resolutions are used. This is because wrong assignments of the material type to voids near material interfaces result in weakly connected or disconnected parts in the inferred structure, making the structure too flexible or statically unstable. This structural disconnection becomes more probable in a higher-resolution mesh as the number of elements increases while less probable with more material types. Therefore, this test clearly shows both the strength and the limitation of element-wise transfer. Global material composition and load paths can be preserved, whereas thin or disconnected members due to material type errors at the material interfaces remain the dominant source of mechanical sensitivity.

\textbf{Three-dimensional MMTO} We further apply the trained XQNN to three-dimensional TO problems without new training. Here, we use a $20\times20\times10$ mesh with four different boundary/loading conditions, $\mathit{3D}_1$--$\mathit{3D}_4$, to assess its generalization performance in a three-dimensional setting (Figure~\ref{fig:prediction}c). The parameters learned from two-dimensional training data are applied voxel by voxel to the descriptors after a three-dimensional MMA run is stopped at $k_{\rm pd}$. XQNN reproduces the main three-dimensional load transfer regions without additional three-dimensional labels or a new network architecture. The representative $\mathit{3D}_2$ predictions show this trend for all material settings. The dominant three-dimensional and internal load paths are reproduced well for all cases, while the three- and four-material predictions lose finer material details around internal interfaces unlike the clean two-material prediction. Across the full $\mathit{3D}_1$--$\mathit{3D}_4$ sets (Figures~S6--S8), the mean accuracy decreases from 89.3\% to 79.6\% and 73.6\% as the number of material types increases. Interestingly, the corresponding mean normalized compliance values are 0.96, 0.98, and 0.83, respectively, indicating that the inferred three-dimensional structures are often stiffer than the GT references despite lower voxel-level accuracies. This is probably because the three-dimensional MMA references can be more prone to local minima than the two-dimensional training solutions, and the learned two-dimensional regularities may sometimes guide XQNN toward mechanically stiffer three-dimensional layouts. However, this effect should be interpreted together with the current limitation of XQNN in that it predicts material type labels element-wise and does not explicitly enforce the global volume constraints during inference.

\subsection{Observable-level explanation and faithfulness test}

The explainable QNN analysis is grounded in the final readout equation, which is linear in the measured observables. For element $e$, the margin between the winning class $c_1$ and the nearest competing class $c_2$ is
\begin{equation}
\Delta z_e=z_{e,c_1}-z_{e,c_2}
=(b_{c_1}-b_{c_2})+\sum_{j=0}^{9}Q_{e,j}\left(W_{j,c_1}-W_{j,c_2}\right).
\label{eq:margin_decomposition_results}
\end{equation}
Conditioned on the trained PQC output $\mathbf{Q}_e$, the readout layer margin is therefore exactly decomposed across observable channels. We define the signed channel contribution and non-negative quantum observable attribution map (QOAM) magnitude as
\begin{equation}
m_{e,j}=Q_{e,j}\left(W_{j,c_1}-W_{j,c_2}\right),
\qquad
a_{e,j}=|m_{e,j}|,
\qquad
A_e=\sum_{j=0}^{9}a_{e,j}.
\label{eq:qoam_results}
\end{equation}
The displayed margin QOAM map is the normalized spatial field $A_e$. High QOAM therefore has a precise meaning: \textit{the measured observables at that element make a large contribution to the local class margin}.

\begin{figure}[p]
\centering
\includegraphics[width=\linewidth, trim={0.0cm 10.5cm 0.0cm 0.0cm}, clip]{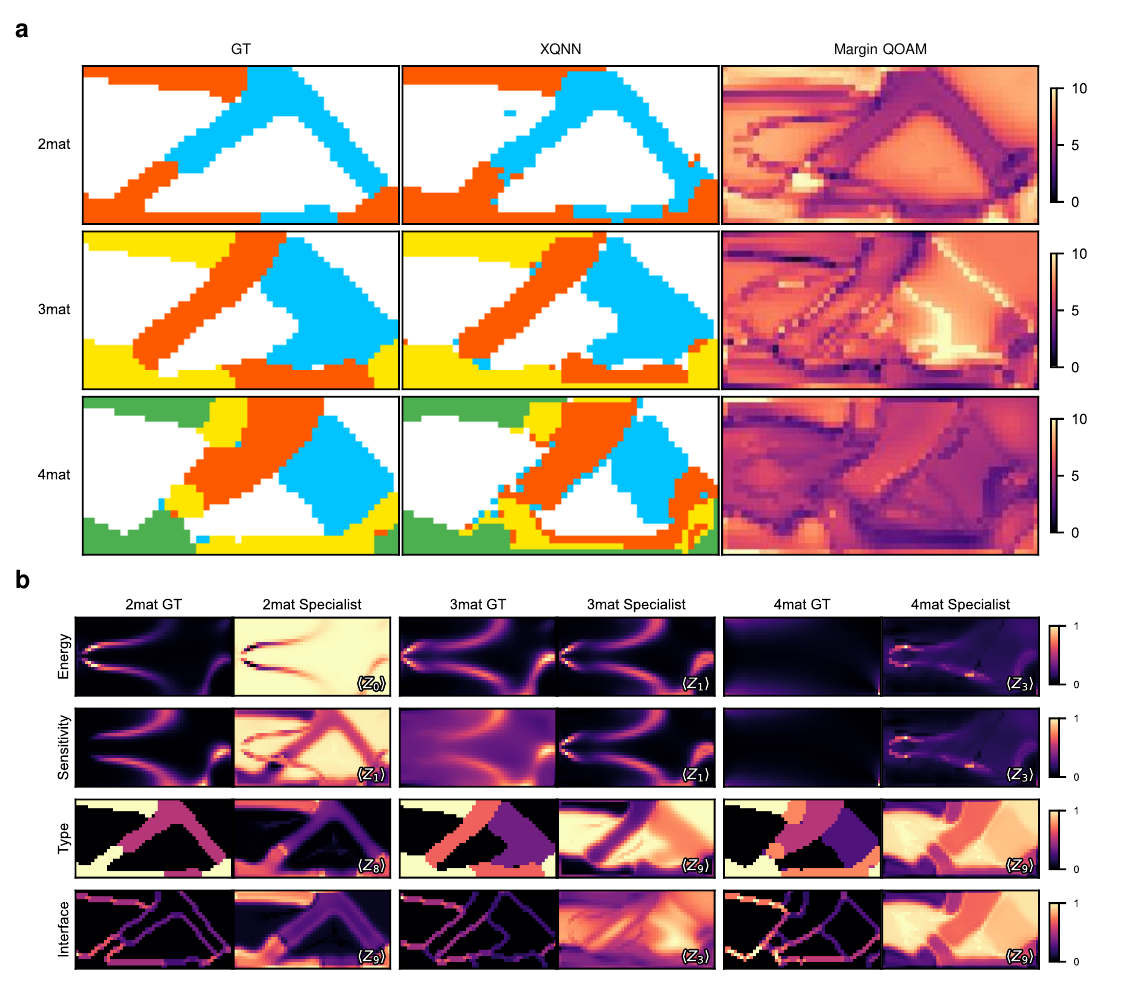}
\caption{\textbf{QOAM and mechanics-facing feature alignment.} \textbf{a}, GT fields, XQNN predictions, and margin QOAM maps for two-, three-, and four-material cases. Margin QOAM decomposes the readout layer class margin into observable channel contributions and highlights regions where measured quantum observables affect the local material decision. \textbf{b}, Validation-selected specialist observables aligned with mechanics-derived target fields, including strain energy, sensitivity, material type label, and interface strength. Strong single channel correlations indicate localized specialist observables, whereas weaker correlations for more complex or interface dominated targets indicate distributed circuit-level representations.}
\label{fig:qoam}
\end{figure}

The attribution maps concentrate on mechanically relevant and decision-sensitive regions (Figure~\ref{fig:qoam}a). High-QOAM regions appear along load-carrying members, material-transition zones, and interface adjacent regions where local material decisions strongly affect the class margin. The two-material case produces relatively localized attributions along the main load path and its boundaries, whereas the three- and four-material cases produce more distributed patterns. This broadening is mechanically plausible because additional material types introduce more material--material interfaces and more local label competition.

Specialist-observable analysis connects individual measured channels to mechanics-derived fields (Figure~\ref{fig:qoam}b). Four targets are used: strain energy, negative sensitivity, final material type label, and interface strength. Here, the interface strength denotes the local finite-difference magnitude of the discrete material type label, used as a proxy for material-boundary intensity rather than as a physical interface length. For each target, the observable channel with the largest absolute Pearson correlation in the validation set is selected and visualized on a representative OOD case. The selected channels and correlation coefficients are summarized in Table~\ref{tab:pearson}. In the two- and three-material cases, several observables behave as localized specialist detectors: strain energy, sensitivity, and material type labels can be associated with individual channels with high values of $|R|$. This provides a direct mechanics-facing interpretation of part of the learned representation.

The same table also shows why a low single channel Pearson correlation should not be read as a failure of the quantum representation. As the number of material types increases, or when the target is the highly local and nonlinear field of interface strength, the largest individual $|R|$ values become smaller. This trend suggests a transition from localized to distributed representation. The ZFeatureMap and CNOT-entangling RealAmplitudes ansatz project and mix the input descriptors across the multi-qubit state before measurement, so information about one mechanical quantity need not remain confined to one same index qubit. The final linear readout can then combine collectively entangled observable responses to recover nonlinear material type and interface relationships. The specialist maps are therefore not intended to be perfect reconstructions of the targets. They test which single observables remain visibly aligned with mechanics, while QOAM and perturbation tests examine how the full observable dictionary supports the decision.

\begin{table}[t]
\centering
\caption{Validation-selected specialist observables and Pearson correlation coefficients. For each mechanics-derived target, the table reports the observable with the largest absolute Pearson correlation on the unified validation set.}
\label{tab:pearson}
\small
\setlength{\tabcolsep}{3pt}
\renewcommand{\arraystretch}{1.12}
\begin{tabular}{@{}p{0.22\linewidth}p{0.23\linewidth}p{0.23\linewidth}p{0.23\linewidth}@{}}
\toprule
Target & 2mat & 3mat & 4mat \\
\midrule
Strain energy & $\langle Z_0\rangle$ ($R=-0.833$) & $\langle Z_1\rangle$ ($R=0.899$) & $\langle Z_3\rangle$ ($R=-0.554$) \\
Sensitivity & $\langle Z_1\rangle$ ($R=-0.603$) & $\langle Z_1\rangle$ ($R=0.844$) & $\langle Z_3\rangle$ ($R=-0.518$) \\
Material type label & $\langle Z_8\rangle$ ($R=0.938$) & $\langle Z_9\rangle$ ($R=-0.881$) & $\langle Z_9\rangle$ ($R=-0.847$) \\
Interface strength & $\langle Z_9\rangle$ ($R=0.345$) & $\langle Z_3\rangle$ ($R=-0.186$) & $\langle Z_9\rangle$ ($R=-0.300$) \\
\bottomrule
\end{tabular}
\end{table}

\begin{figure}[p]
\centering
\includegraphics[width=\linewidth, trim={0.0cm 12.0cm 0.0cm 0.0cm}, clip]{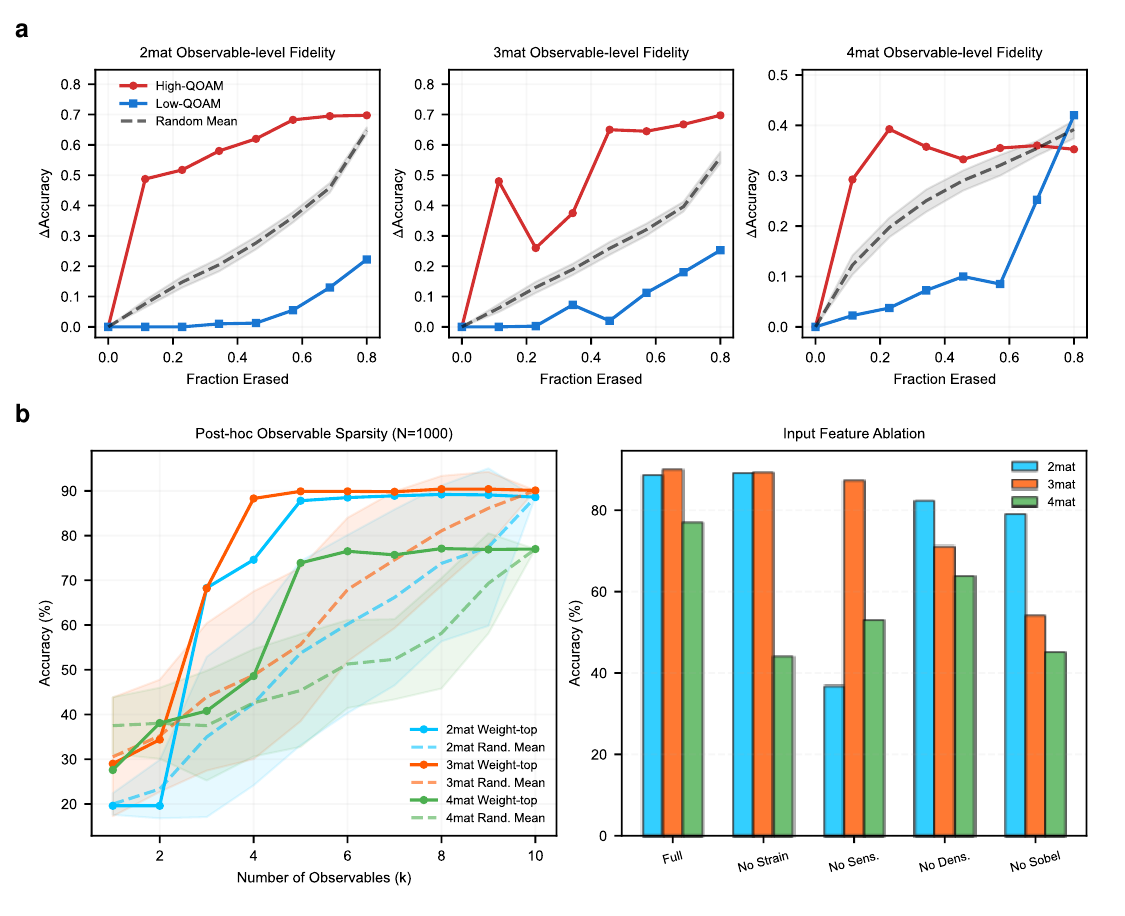}
\caption{\textbf{Observable-level faithfulness, sparsity, and input feature ablation.} \textbf{a}, Observable-level erasure response under high-QOAM, low-QOAM, and random channel masking. \textbf{b}, Post-hoc sparsity obtained by retaining only the highest-ranked observable channels; plateau regions indicate that the dominant margin-carrying observables have already been retained. The right panel shows input feature ablation obtained by removing strain energy, sensitivity, density, or the Sobel boundary hint.}
\label{fig:ablation}
\end{figure}

Faithfulness is tested by perturbing observables and inputs after training (Figure~\ref{fig:ablation}). Observable erasure replaces selected channels by neutral-reference values computed from a constant mid-range input vector. Masking high-QOAM channels produces a stronger erasure response than masking low-QOAM channels, with random masking between the two targeted cases (Figure~\ref{fig:ablation}a). This ordering is important because it tests the explanation, not only the prediction. Channels identified by the QOAM margin decomposition are the channels whose removal most strongly changes the classifier response. QOAM therefore highlights decision-relevant observables rather than visually plausible but unused features.

Post-hoc sparsity and input feature ablation provide complementary checks (Figure~\ref{fig:ablation}b). When only a small number of observables are retained, the channels ranked highly by readout-weight magnitude outperform random channel subsets, showing that the trained observable dictionary contains a compact decision core. The sparsity curves also explain why the response can plateau after several top-ranked channels have been retained: once the dominant margin-carrying observables are present, additional low-ranked channels mainly add redundant or case-specific information and may not change the predicted class for many elements. The four-material task requires a broader subset than the two- and three-material tasks, which is consistent with the larger number of material types and material--material interfaces that must be separated. Input-feature ablation further shows that strain energy, sensitivity, density, and the Sobel boundary hint play different roles. Energy and sensitivity support load-path recognition, density carries the evolving material layout, and the Sobel boundary hint contributes interface sensitive information. No single descriptor group explains all predictions, and the limited effect of removing some groups in selected cases is consistent with distributed encoding, where related mechanical information can be partially recovered from other observables. Together with the validation set specialist maps, these perturbation tests make the explanations causal at the readout level and mechanically interpretable at the input feature level.

\section{Discussion}

The results support an interpretation of XQNN: it is an observable-based surrogate for MMTO, not a black-box quantum classifier. The MMA solver supplies mechanics-derived element descriptors, the ZFeatureMap writes these descriptors into quantum phase-encoded qubit states, the RealAmplitudes ansatz transforms those states, and the readout layer uses measured $Z$ observables to predict material type labels. Retaining qubit-wise observables is central to the method because the same quantities used for prediction can also be decomposed by QOAM, correlated with mechanics-derived targets, erased, compressed, and input-ablated.

Three conclusions follow from this design. First, a compact ten-qubit model can capture structural information that transfers across OOD two-dimensional boundary/loading conditions and mesh resolutions, and it can be applied to voxelized three-dimensional domains without additional three-dimensional training. Transfer is imperfect, especially in four-material cases, but the dominant load-carrying topology is usually preserved while errors concentrate near interfaces and thin material regions. The accuracy--compliance pairs are important here: pixel-wise material type accuracy can remain high even when a small connectivity error strongly affects compliance, so both metrics are needed to evaluate a mechanics surrogate. Second, retaining qubit-wise observables provides an auditable intermediate representation rather than a single compressed scalar response. The separate measured $Z$ channels preserve independent readout degrees of freedom generated after superposition, phase encoding, and entangling transformations, allowing the classifier and the explanation analyses to operate on the same measurable variables. Third, the Sobel boundary hint contributes a complementary signal rather than a complete explanation. It gives the tenth qubit local edge-sensitive information, while the ablation results show that strain energy, sensitivity, density, and the Sobel boundary hint are all used, with their relative importance changing across the number of material types.

The specialist observable results clarify how the quantum circuit should be interpreted. A tempting but incorrect reading would be to associate each measured qubit directly with the same index input feature. The validation correlations show a different picture: after quantum phase encoding and CNOT-mediated mixing, observables become circuit-level responses to the descriptor vector. Strong single channel correlations in simpler cases provide localized interpretability, whereas lower single channel correlations for the four-material and interface-strength targets indicate that the relevant information is increasingly distributed across the observable dictionary. This distributed representation is useful because the readout can reconstruct complex material types and interface decisions from collective observable responses, but it also requires explanation methods that operate on the measured observables rather than on the raw inputs alone. QOAM, specialist alignment, and perturbation tests therefore play complementary roles: QOAM decomposes the decision margin, specialist alignment relates observables to mechanics-derived targets, and erasure/sparsity tests whether the identified channels are actually used by the classifier.

The main limitation of the current work is that all quantum evaluations use exact statevector simulation. Real devices would introduce shot noise, gate errors, and decoherence, and hence, it should be tested whether the observable-level explanations remain stable under those effects \cite{CerezoVerdonColes2022}. Nevertheless, the proposed framework is not limited to MMTO only. Any mechanics surrogate that converts local descriptors into point-wise or element-wise labels can use the same design principle: retain a compact dictionary of measured observables, train a readout layer, and interrogate the resulting decisions with margin decomposition, specialist alignment, and perturbation tests.

\section{Methods}

\subsection{MMTO simulations and dataset organization}

Ground-truth (GT) layouts are generated with an MMA-based multi-material topology optimization solver adapted from the MATLAB code of Zheng \textit{et al.}, which provides complete two-dimensional and three-dimensional implementations for compliance minimization of multi-material continuum structures using mapping-based interpolation, density filtering, projection, and sensitivity analysis \cite{ZhengEtAl2024MMTOCode}. In our dataset generation, the optimization minimizes compliance subject to finite element equilibrium, box constraints, and one volume constraint per material type. The material type settings, relative Young's moduli, and volume constraints used for the main two-dimensional datasets are summarized in Table~S2. The terminal GT/reference layouts are generated by running MMA until the grayscale ratio falls below 0.02 for the regular two-dimensional and three-dimensional simulations and below 0.01 for the high-resolution two-dimensional simulations. These terminal criteria define the converged reference layouts, whereas $G_{\rm target}$ defines the earlier prediction decision state used for XQNN inference. Details of the material interpolation, finite element discretization, filtering, and three-dimensional problem settings are given in Supplementary Note~1.
We use the disjoint training, validation, and final evaluation roles defined in Results and Figure~\ref{fig:training}--\ref{fig:prediction}. The same split is used throughout XQNN training and evaluation, and the boundary/loading definitions and mesh sizes are summarized in Table~S1. This Methods section therefore specifies how the split is used in simulation, preprocessing, and training rather than repeating the full problem definitions.

\subsection{Preprocessing and descriptor construction}

The terminal MMTO material type label fields are refined with a $3\times3$ mode filter, implemented as majority voting over the local label neighborhood, to suppress isolated high-frequency numerical artifacts and checkerboard-like patterns along material type boundaries. The input descriptor for each element consists of strain energy, sensitivity, and density sampled at $k_{\rm em}$, $k_{\rm em}+5$, and $k_{\rm pd}$, together with a Sobel boundary hint evaluated from the prediction stage density field. The emergence iteration $k_{\rm em}$ is selected using $G_{\rm em}=0.5$, and $k_{\rm pd}$ is selected using $G_{\rm target}=0.20$, 0.15, and 0.10 for the two-, three-, and four-material cases, respectively. These $G_{\rm target}$ values are grayscale-ratio thresholds and are not used as a synonym for the number of material types.

For each two-, three-, or four-material case, min--max scaling to $[0,1]$ is fitted independently and only on the corresponding training designs. The fitted transform is reused unchanged for validation, OOD, high-resolution, and three-dimensional inference. Scaled descriptors are clipped to $[0,1]$ and then mapped to quantum feature map angles in $[0,\pi]$. This training-only scaling prevents data leakage from evaluation distributions while preserving the valid input range of the ZFeatureMap. Detailed descriptor equations and Sobel definitions are provided in Supplementary Note~2.

\subsection{Model training and analysis}

The XQNN circuit uses ten qubits, a ZFeatureMap with REPS=2, and a RealAmplitudes ansatz with REPS=3. The ZFeatureMap uses Hadamard and phase gates, and the RealAmplitudes ansatz uses trainable $R_Y$ rotations with CNOT entanglers, as shown in Figure~\ref{fig:workflow}b,c. Exact statevector simulation is used for all main results, and the measured qubit-wise $Z$ observables are passed to a linear softmax readout. Training minimizes class balanced multi-class cross-entropy with Adam optimization. At each epoch, class balanced random sampling draws up to 20 elements per class from each of the three training designs, giving 180, 240, and 300 sampled elements per epoch for the two-, three-, and four-material tasks, respectively. The model with the highest mean validation accuracy on the full $V_1$--$V_3$ validation set is retained. Details of the training schedule, parameter shift gradients, sampling protocol, and configuration studies are provided in Supplementary Note~3.

All quantum evaluations use exact statevector simulation in Python 3.12.10 with Qiskit SDK v2.3.1 and Qiskit Aer v0.17.2 on an AMD Ryzen 7 9700X 8-Core Processor. The QOAM, specialist observable, erasure, sparsity, and input feature ablation protocols are summarized in Results and detailed in Supplementary Note~4.

\section{Data availability}
The data that support the plots within this paper and other findings of
this study are available from the corresponding author upon reasonable
request.

\section{Code availability}
The computer codes used to generate the results of this paper will be made available on request.

\section{Acknowledgements}
This work was supported by the National Research Foundation of Korea (NRF) grant funded by the Ministry of Science and ICT (RS-2024-00346176) and the Development of Core Technologies for Manufacturing Foundation Models Program (RS-2025-25458052) funded by the Ministry of Trade, Industry \& Energy (MOTIE, Korea). D.J. also thanks Minwoo Kim at Seoul National
University for fruitful discussions and helpful comments.

\section{Author Contributions}
D.J. conceived the project, designed the research, proposed the algorithm,
and assembled input data; D.J. and D.-N.K. analyzed output data and wrote the initial manuscript; N.S. and K.T. provided the prior research code, guidance on the test direction, and result validation; D.-N.K. supervised the project;
all authors contributed to the writing of the manuscript.

\section{Competing Interests}
The authors declare no competing interests.

\bibliography{sn-bibliography}

\clearpage

\clearpage
\clearpage
\setcounter{tocdepth}{2}
\renewcommand{\contentsname}{Contents}
\renewcommand{\theequation}{S\arabic{equation}}
\renewcommand{\thefigure}{S\arabic{figure}}
\renewcommand{\thetable}{S\arabic{table}}
\renewcommand{\figurename}{Figure}
\renewcommand{\tablename}{Table}
\setcounter{equation}{0}
\setcounter{figure}{0}
\setcounter{table}{0}

\hypersetup{pageanchor=false,linkcolor=black,citecolor=black,urlcolor=black}
\providecommand{\theHpage}{}
\providecommand{\theHequation}{}
\providecommand{\theHfigure}{}
\providecommand{\theHtable}{}
\renewcommand{\theHpage}{S\arabic{page}}
\renewcommand{\theHequation}{S\arabic{equation}}
\renewcommand{\theHfigure}{S\arabic{figure}}
\renewcommand{\theHtable}{S\arabic{table}}
\captionsetup{font=small,labelfont=bf,labelsep=period}
\newcolumntype{Y}{>{\raggedright\arraybackslash}X}
\makeatletter
\renewcommand{\tableofcontents}{%
  \section*{\contentsname}%
  \@starttoc{stoc}%
}
\let\origaddcontentsline\addcontentsline
\renewcommand{\addcontentsline}[3]{%
  \def\supp@tocarg{#1}%
  \def\supp@tocname{toc}%
  \ifx\supp@tocarg\supp@tocname
    \origaddcontentsline{stoc}{#2}{#3}%
  \else
    \origaddcontentsline{#1}{#2}{#3}%
  \fi
}
\renewenvironment{thebibliography}[1]
{%
  \setcounter{NAT@ctr}{0}%
  \list{\@biblabel{\@arabic\c@enumiv}}%
  {%
    \settowidth\labelwidth{\@biblabel{#1}}%
    \leftmargin\labelwidth
    \advance\leftmargin\labelsep
    \usecounter{enumiv}%
    \let\p@enumiv\@empty
    \renewcommand\theenumiv{\@arabic\c@enumiv}%
  }%
  \sloppy
  \clubpenalty4000
  \@clubpenalty \clubpenalty
  \widowpenalty4000%
  \sfcode`\.\@m
}
{%
  \def\@noitemerr{\@latex@warning{Empty `thebibliography' environment}}%
  \endlist
}
\makeatother

\begin{titlepage}
\centering
\vspace*{2.0cm}
{\Large\bfseries Supplementary Information\par}
\vspace{1.5cm}
{\LARGE\bfseries Explainable quantum neural networks for multi-material topology optimization\par}
\vspace{1.5cm}
{\large Dahyun Joo$^{1}$, Naruethep Sukulthanasorn$^{2}$, Kenjiro Terada$^{2,3}$, and Do-Nyun Kim$^{1,4,5,*}$\par}
\vspace{1.0cm}
\begin{flushleft}
$^{1}$Department of Mechanical Engineering, Seoul National University, Seoul 08826, Republic of Korea\\
$^{2}$International Research Institute of Disaster Science, Tohoku University, Sendai 980-8572, Japan\\
$^{3}$Department of Civil and Environmental Engineering, Tohoku University, Sendai 980-8579, Japan\\
$^{4}$Institute of Advanced Machines and Design, Seoul National University, Seoul 08826, Republic of Korea\\
$^{5}$Institute of Engineering Research, Seoul National University, Seoul 08826, Republic of Korea\\
$^{*}$Correspondence: dnkim@snu.ac.kr
\end{flushleft}
\vfill
\end{titlepage}
\clearpage

\pagenumbering{arabic}
\setcounter{page}{1}

\tableofcontents
\clearpage

\phantomsection
\addcontentsline{toc}{section}{Supplementary Notes}
\section*{Supplementary Notes}

\phantomsection
\addcontentsline{toc}{subsection}{Supplementary Note 1. MMTO ground-truth simulation and material interpolation}
\subsection*{Supplementary Note 1. MMTO ground-truth simulation and material interpolation}

Ground-truth (GT) material layouts are generated with a multi-material topology optimization (MMTO) solver based on the method of moving asymptotes (MMA). MMA is an iterative gradient-based optimization method that replaces the original nonlinear structural optimization problem with a sequence of convex approximate subproblems \cite{Svanberg1987}. Starting from an initial material distribution, each MMA iteration performs finite element analysis, evaluates compliance and sensitivities, updates the material design variables under volume constraints, and repeats this process until a sufficiently discrete material layout is obtained.

The solver is adapted from the MATLAB code of Zheng \textit{et al.}, which provides complete two-dimensional and three-dimensional implementations for compliance minimization of multi-material continuum structures using mapping-based interpolation, density filtering, Heaviside projection, and sensitivity analysis \cite{Zheng2024}. In this workflow, the MMTO simulation produces three types of information used by the explainable quantum neural network (XQNN): intermediate strain energy fields, intermediate sensitivity fields, and intermediate density fields at selected MMA iterations. The terminal material type label field obtained at the end of the MMTO simulation is used as the GT target after the $3\times3$ mode filtering step described in Supplementary Note~2. Thus, the same MMTO history provides both the input descriptors and the supervised material type labels.

The two-dimensional datasets use four-node quadrilateral plane-stress elements with Poisson's ratio $\nu=0.3$. The three-dimensional datasets use eight-node hexahedral elements with $2\times2\times2$ Gauss integration. Density filtering and Heaviside projection are applied before stiffness assembly. For terminal GT/reference generation, the regular two-dimensional and three-dimensional MMA simulations are stopped when the grayscale ratio satisfies $G<0.02$, whereas the high-resolution two-dimensional simulations are stopped at $G<0.01$. These terminal criteria are distinct from $G_{\rm target}$, which defines the earlier prediction decision state used as the XQNN input. The dataset roles are summarized in Table~\ref{tab:s_dataset_roles}, and the main two-dimensional material settings are summarized in Table~\ref{tab:s_dataset_settings}. In all material layout plots, void is shown in white; the first, second, third, and fourth solid material types in the ordered stiffness list are shown in blue, orange, yellow, and green, respectively. Thus, the two-material layouts use blue and orange, the three-material layouts additionally use yellow, and the four-material layouts additionally use green.

For a design domain discretized into $N_e$ finite elements and $M$ solid material types, the design variable $\rho_{e,m}\in[0,1]$ denotes the amount of material type $m$ assigned to element $e$. The optimization minimizes compliance,
\begin{equation}
c(\bm{\rho})=\mathbf{F}^T\mathbf{U},
\label{eq:s_compliance}
\end{equation}
subject to equilibrium,
\begin{equation}
\mathbf{K}(\bm{\rho})\mathbf{U}=\mathbf{F},
\label{eq:s_equilibrium}
\end{equation}
box constraints and one volume constraint per material type,
\begin{equation}
0\leq \rho_{e,m}\leq1,
\qquad
\frac{1}{N_e}\sum_{e=1}^{N_e}\rho^{\rm phys}_{e,m}\leq V_m,
\qquad m=1,\ldots,M .
\label{eq:s_volume}
\end{equation}

The material interpolation treats material types symmetrically. With $p_{\rm norm}=6$ and $\delta=10^{-9}$,
\begin{equation}
S_e=\sum_{m=1}^{M}\left(\rho^{\rm phys}_{e,m}\right)^{p_{\rm norm}}+\delta,
\qquad
\widehat{\rho}_{e,m}=\frac{\rho^{\rm phys}_{e,m}S_e^{1/p_{\rm norm}}}{\sum_{j=1}^{M}\rho^{\rm phys}_{e,j}+\delta}.
\label{eq:s_rhohat}
\end{equation}
The effective Young's modulus is
\begin{equation}
E_e=E_{\min}+\sum_{m=1}^{M}E_m\widehat{\rho}_{e,m}^{p_{\rm penal}},
\qquad p_{\rm penal}=3,
\label{eq:s_effective_modulus}
\end{equation}
where $E_{\min}=10^{-9}$ represents void.

At each MMA iteration, the filtered and projected densities define the element stiffness matrix through Eq.~\eqref{eq:s_effective_modulus}. The global stiffness matrix is assembled, the equilibrium equation in Eq.~\eqref{eq:s_equilibrium} is solved, and the compliance in Eq.~\eqref{eq:s_compliance} is evaluated. The strain energy descriptor used by XQNN is obtained from the element-wise contribution to this finite element solution, and the sensitivity descriptor is obtained from the compliance derivative with respect to the local material variables. The final material type label is assigned from the terminal projected densities by selecting the dominant material type at each element, with void represented by the additional readout class when no solid material type dominates. The mode filter is then applied only to the terminal label field, not to the finite element analysis itself.

\phantomsection
\addcontentsline{toc}{subsection}{Supplementary Note 2. Dataset roles, selected iterations, preprocessing, and scaling}
\subsection*{Supplementary Note 2. Dataset roles, selected iterations, preprocessing, and scaling}

The training, validation, and OOD test roles are kept disjoint throughout the main article and this Supplementary Information (Table~\ref{tab:s_dataset_roles}). Training problems $T_1$--$T_3$ are used to fit the circuit and readout parameters. Validation problems $V_1$--$V_3$ are used only for model selection and early stopping. OOD problems $O_1$--$O_4$ are evaluated only after model selection. High-resolution transfer cases $\mathit{HR}_1$--$\mathit{HR}_3$ and three-dimensional transfer cases $\mathit{3D}_1$--$\mathit{3D}_4$ are also excluded from training and validation. Figure~2 of the main text defines the training and validation boundary/loading conditions and shows the selected input fields and convergence behavior. Figure~3 of the main text separates final generalization into OOD, high-resolution, and three-dimensional transfer subpanels. Figure~3a reports material type accuracy together with normalized compliance, defined as $C_{\rm pred}/C_{\rm GT}$, for the OOD cases; Figure~3b shows the high-resolution transfer layouts; and Figure~3c shows representative three-dimensional transfer results. The full GT--prediction layout pairs for all OOD, high-resolution, and three-dimensional cases are provided in Figures~\ref{fig:full2mat}--\ref{fig:full4mat}, while Table~\ref{tab:s_dataset_roles} summarizes their roles, mesh sizes, and boundary/loading descriptions.

The grayscale ratio
\begin{equation}
G(k)=\frac{1}{N_eM}\sum_{e=1}^{N_e}\sum_{m=1}^{M}4\rho^{\rm phys}_{e,m}(k)\{1-\rho^{\rm phys}_{e,m}(k)\}
\label{eq:s_grayscale}
\end{equation}
quantifies the discreteness of the evolving material field at MMA iteration $k$. The emergence iteration $k_{\rm em}$ is selected using $G_{\rm em}=0.5$, which marks the stage at which a recognizable but still gray structural layout has emerged. The additional $k_{\rm em}+5$ iteration retains a slightly more mature post-emergence state while still remaining earlier than the final optimized design. The prediction decision iteration $k_{\rm pd}$ is selected using $G_{\rm target}$, which determines when the early stopped optimization state is converted into the XQNN input. In the main two-, three-, and four-material cases, $G_{\rm target}=0.20$, 0.15, and 0.10 are used, respectively. These values are grayscale-ratio thresholds and are not used as a synonym for the number of material types. Lowering $G_{\rm target}$ moves the prediction decision state closer to the final optimized layout, making the GT labels sharper and usually improving validation accuracy. This improvement comes at the cost of a smaller acceleration benefit, because more MMA iterations must be completed before XQNN inference begins. The selected values therefore represent a time--accuracy compromise: $G_{\rm target}=0.20$ for two materials, 0.15 for three materials, and 0.10 for four materials. The corresponding validation tradeoff is evaluated in the configuration study described below.

Each selected MMA iteration contributes strain energy, sensitivity, and density descriptors. The strain energy descriptor is computed from the element displacement vector $\mathbf{u}_e(k)$ and element stiffness matrix $\mathbf{K}_e(k)$ as
\begin{equation}
E_e(k)=\mathbf{u}_e(k)^T\mathbf{K}_e(k)\mathbf{u}_e(k),
\label{eq:s_strain_energy}
\end{equation}
up to the common constant convention used in finite element strain energy definitions. The sensitivity descriptor is
\begin{equation}
\frac{\partial c}{\partial \rho_{e,m}}=-\mathbf{u}_e(k)^T
\frac{\partial \mathbf{K}_e(k)}{\partial \rho_{e,m}}\mathbf{u}_e(k),
\qquad
S_e(k)=\sum_{m=1}^{M}\left|\frac{\partial c}{\partial \rho_{e,m}}\right|,
\label{eq:s_sensitivity}
\end{equation}
where the chain rule includes filtering, projection, and material interpolation. The density descriptor $D_e(k)$ records the projected density field used by the optimizer.

The terminal material type label fields are refined before learning by applying a $3\times3$ mode filter, implemented as majority voting over the local label neighborhood. This filter suppresses isolated high-frequency numerical artifacts and checkerboard-like patterns along material type boundaries without changing the finite element optimization procedure used to produce the reference layouts. The filtered label field is used consistently for training labels, validation labels, and reported material type accuracy calculations.

The Sobel boundary hint is evaluated from the prediction stage density field. In two dimensions,
\begin{equation}
B_e(k_{\rm pd})=\sqrt{\{(G_x*D(k_{\rm pd}))_e\}^2+\{(G_y*D(k_{\rm pd}))_e\}^2},
\label{eq:s_sobel}
\end{equation}
where $G_x$ and $G_y$ denote the horizontal and vertical Sobel kernels, and $*$ denotes convolution on the element mesh. The three-dimensional implementation is designed to keep the filter scale and apparent boundary thickness close to the two-dimensional case while limiting mesh blurring. For a derivative along a target axis, we first apply the central finite-difference stencil $[-1,0,1]$ along that axis and then apply the smoothing kernel $[1,2,1]/2$ sequentially along the other two axes. This Sobel operation follows a two-dimensional filtering scale while being applied in three dimensions, giving directional components $G_x*D$, $G_y*D$, and $G_z*D$, and the unsharpened gradient magnitude is
\begin{equation}
\bar{B}_e^{\rm 3D}(k_{\rm pd})=\sqrt{\{(G_x*D)_e\}^2+\{(G_y*D)_e\}^2+\{(G_z*D)_e\}^2}.
\label{eq:s_sobel3d_raw}
\end{equation}
After normalizing $\bar{B}_e^{\rm 3D}$ to $[0,1]$, we apply power sharpening to suppress excessive voxel mesh spreading,
\begin{equation}
B_e^{\rm 3D}(k_{\rm pd})=\left\{\operatorname{normalize}\left(\bar{B}_e^{\rm 3D}(k_{\rm pd})\right)\right\}^{2}.
\label{eq:s_sobel3d}
\end{equation}
The resulting three-dimensional Sobel boundary hint is stored as a continuous normalized floating-point field without binarization, preserving gradual boundary variations around material type interfaces. The Sobel descriptor is not used to define the GT label; it is an input feature that exposes local boundary intensity to the learning pipeline.

For each two-, three-, or four-material case, min--max scaling to $[0,1]$ is fitted independently and only on the corresponding three training designs. The fitted scaler defines the transform $\mathcal{S}_{M,q}$, is reused unchanged at validation and test time, and is not refitted on OOD, high-resolution or three-dimensional cases. This prevents data leakage from evaluation distributions. Clipping prevents out-of-range descriptors from producing values outside the intended feature map range and allows the normalized feature to be mapped to the final quantum feature map angle,
\begin{equation}
\tilde{x}_{e,q}=\operatorname{clip}\!\left(\mathcal{S}_{M,q}(x_{e,q}),0,1\right),
\qquad
\alpha_{e,q}=\pi\tilde{x}_{e,q}.
\label{eq:s_scaling}
\end{equation}

\phantomsection
\addcontentsline{toc}{subsection}{Supplementary Note 3. XQNN training and configuration selection}
\subsection*{Supplementary Note 3. XQNN training and configuration selection}

The XQNN circuit uses $n_q=10$ qubits. The normalized element descriptor is encoded with a ZFeatureMap with REPS=2 and transformed by a RealAmplitudes ansatz with REPS=3 and CNOT entanglers. Exact statevector simulation is used for all main and supplementary results using Qiskit workflows \cite{Qiskit}. All quantum computations were carried out in a Python 3.12.10 environment using Qiskit SDK v2.3.1 and Qiskit Aer v0.17.2. Classical post-processing and statistical analyses used NumPy v2.4.4, SciPy v1.17.1, and scikit-learn v1.8.0. The numerical experiments were run on a workstation equipped with an AMD Ryzen 7 9700X 8-Core Processor. For ten qubits, the statevector has $2^{10}=1024$ complex amplitudes, so the observable values can be evaluated directly without finite shot sampling.

For completeness, the circuit notation used in Figure~1 can be written as follows. A one-repetition feature map applies Hadamard and phase gates to the normalized descriptor angles, and the RealAmplitudes block applies trainable $R_Y$ rotations with CNOT entanglers. In compact notation,
\begin{equation}
|\psi_e(\bm{\theta})\rangle=U_{\rm RA}^{(3)}(\bm{\theta})U_{\rm ZF}^{(2)}(\tilde{\mathbf{x}}_e)|0\rangle^{\otimes 10},
\label{eq:s_pqc_state}
\end{equation}
and the qubit-wise observables are
\begin{equation}
\mathbf{Q}_e=(\langle Z_0\rangle_e,\ldots,\langle Z_9\rangle_e)^T,\qquad \langle Z_i\rangle_e=\langle\psi_e|Z_i|\psi_e\rangle .
\label{eq:s_observables}
\end{equation}
The readout then applies the linear softmax layer described in the main text. This notation records the mathematical circuit object, while the main Results section focuses on the Figure~1 workflow and the mechanics meaning of the descriptors.

Training minimizes multi-class cross-entropy with L2 regularization on the readout layer weights,
\begin{equation}
\mathcal{L}=-\log p_{e,y_e}+\frac{\lambda}{2}\|\bm{\theta}_{\rm readout}\|_2^2,
\qquad \lambda=5\times10^{-5}.
\label{eq:s_loss}
\end{equation}
Adam optimization \cite{KingmaBa2015} is used with $\beta_1=0.9$, $\beta_2=0.999$, $\epsilon=10^{-7}$, initial learning rate 0.1 with linear decay to 0, gradient clipping to $[-5.0,5.0]$, and 100 epochs. Circuit gradients are computed by the parameter shift rule with shift $\pm\pi/2$.

Training uses class balanced random sampling with replacement to reduce class imbalance and overfitting. The default training set uses three boundary/loading designs ($D=3$): $T_1$ (left fixed cantilever), $T_2$ (both ends fixed), and $T_3$ (bottom fixed wall). Each design contains 1,250 elements. At every epoch, up to 20 elements per class are randomly sampled from each design. Since the readout has void plus $M$ solid material types, this gives 60 samples per design for two materials, 80 samples per design for three materials, and 100 samples per design for four materials. Across the three training designs, each epoch therefore uses 180, 240, and 300 sampled training elements for the two-, three-, and four-material tasks, respectively. The sampled elements are shuffled at the start of each epoch and used to form mini-batches of size 32.

The training loss curves in the main and supplementary figures are computed from these class balanced sampled training subsets. In contrast, the validation accuracy curves are computed deterministically on the full validation set at the end of each epoch: $V_1$, $V_2$, and $V_3$ each contain 1,250 elements, giving 3,750 validation elements in total. The plotted validation accuracy is the mean accuracy over these three complete validation problems, not an accuracy computed from a sampled subset. The weights with the highest mean validation accuracy are retained as the best model.

\begin{figure}[H]
\centering
\includegraphics[width=\linewidth, trim={0.0cm 12.0cm 0.0cm 0.0cm}, clip]{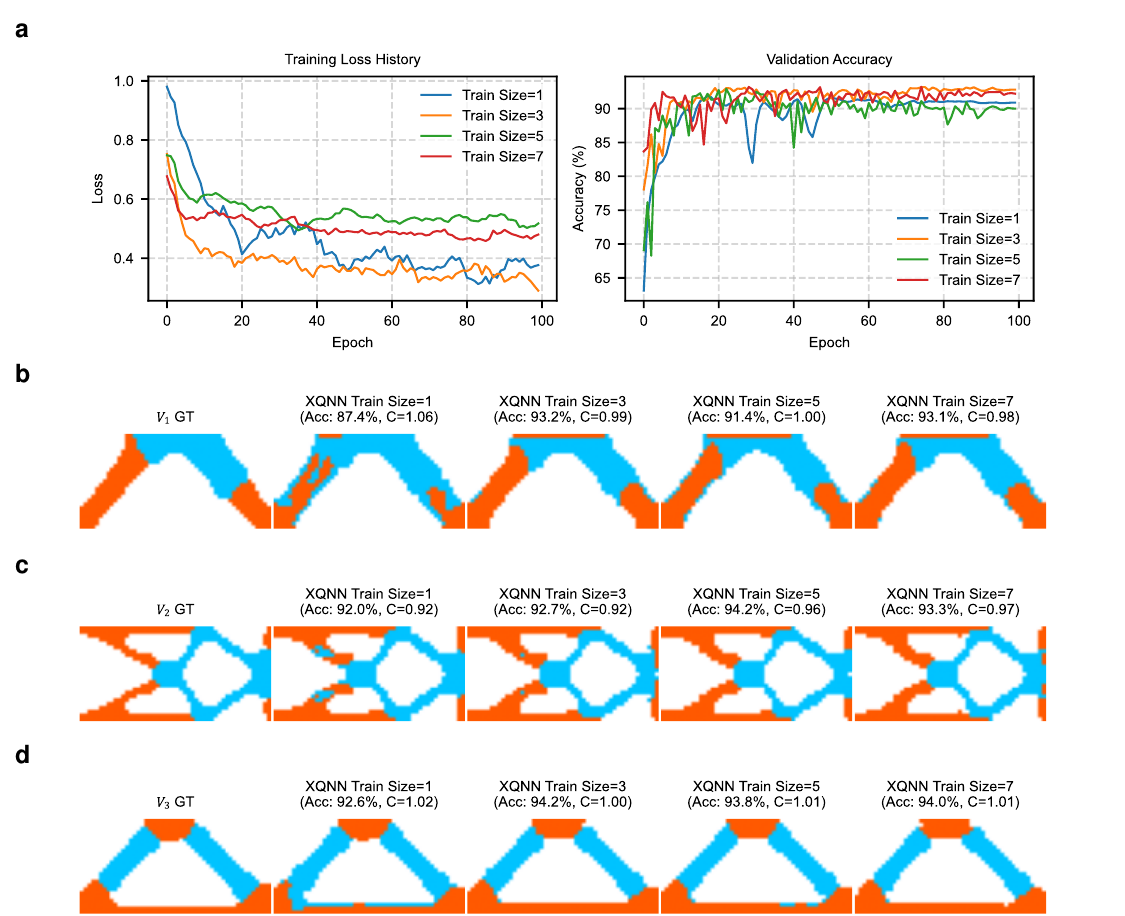}
\caption{\textbf{Sensitivity to number of training designs.} The number of training designs was varied across $D\in\{1,3,5,7\}$: $D=1$ uses $T_1$; $D=3$ uses $T_1$--$T_3$ and is the default setting; $D=5$ adds $O_1$ and $O_2$; and $D=7$ adds $O_1$--$O_4$. \textbf{a}, Training loss and mean validation accuracy over $V_1$--$V_3$. \textbf{b--d}, Validation predictions for $V_1$--$V_3$ compared with the corresponding GT layouts. The default $D=3$ setting gives sufficient validation accuracy without using OOD designs for training, so it is used for the main results.}
\label{fig:datasize}
\addcontentsline{toc}{subsection}{Figure S1. Sensitivity to number of training designs}
\end{figure}

\begin{figure}[H]
\centering
\includegraphics[width=\linewidth, trim={0.0cm 5.5cm 0.0cm 0.0cm}, clip]{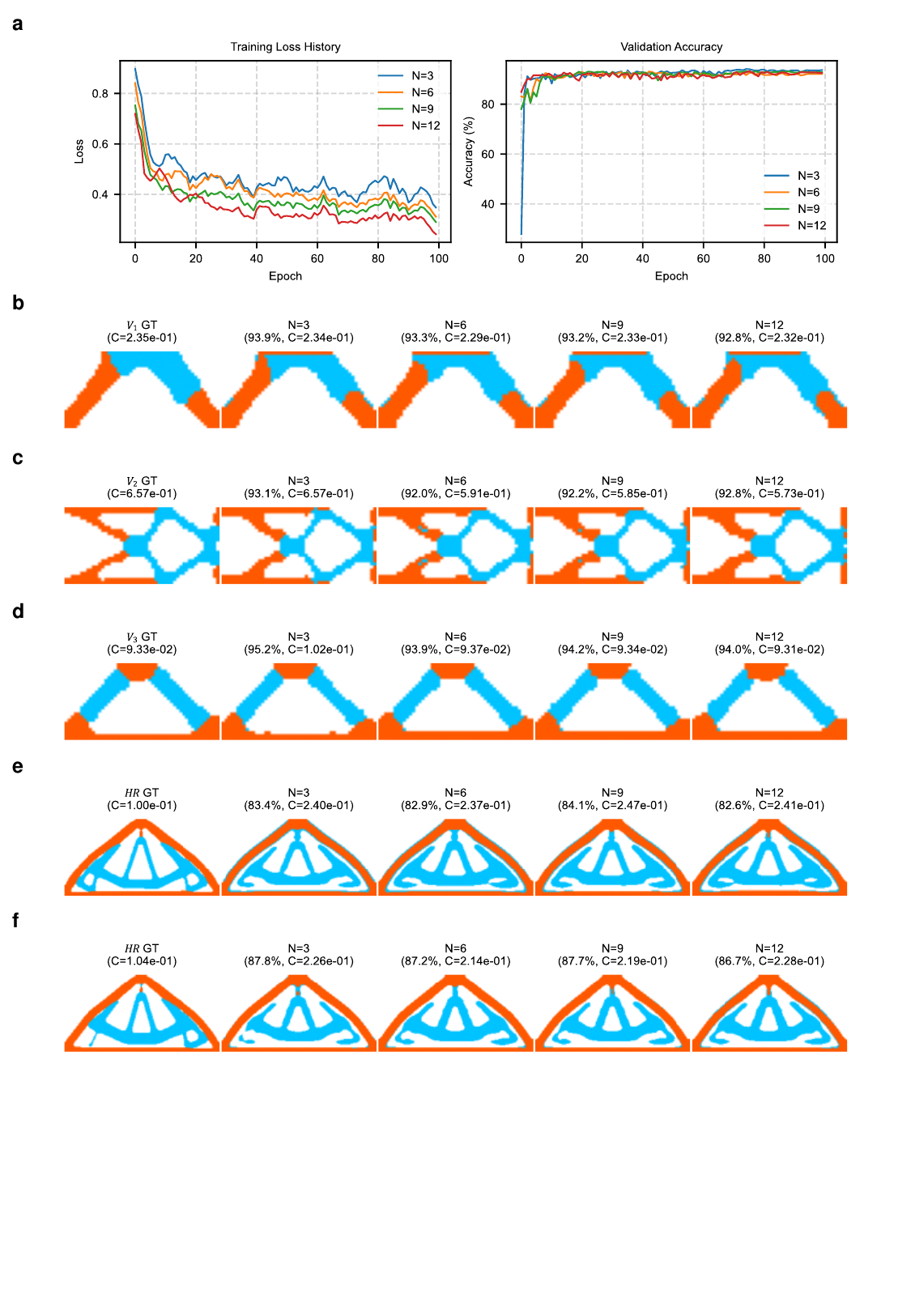}
\caption{\textbf{Effect of optimizer history input size and qubit count.} The number of physical descriptors before appending the Sobel boundary hint was varied as $N\in\{3,6,9,12\}$, corresponding to 4, 7, 10, and 13 qubits after the boundary hint is included. \textbf{a}, Training loss decreases as more optimizer history information is supplied, whereas the mean validation accuracy over $V_1$--$V_3$ changes only modestly. \textbf{b--d}, Validation predictions for $V_1$--$V_3$. \textbf{e,f}, High-resolution $\mathit{HR}_3$ transfer tests under two additional two-material stiffness settings, $[1,4]$ and $[2,11]$, respectively; blue and orange denote the lower- and higher-stiffness solid material types. With limited input history, the upper orange connector and thin blue members are less stable or disconnected, whereas multiple MMA milestones better preserve connected material paths and hole shapes. The default setting is $N=9$, corresponding to $k_{\rm em}$, $k_{\rm em}+5$, and $k_{\rm pd}$, plus the Sobel boundary hint.}
\label{fig:numqbit}
\addcontentsline{toc}{subsection}{Figure S2. Effect of optimizer history input size and qubit count}
\end{figure}

\begin{figure}[H]
\centering
\includegraphics[width=\linewidth, trim={0.0cm 12.0cm 0.0cm 0.0cm}, clip]{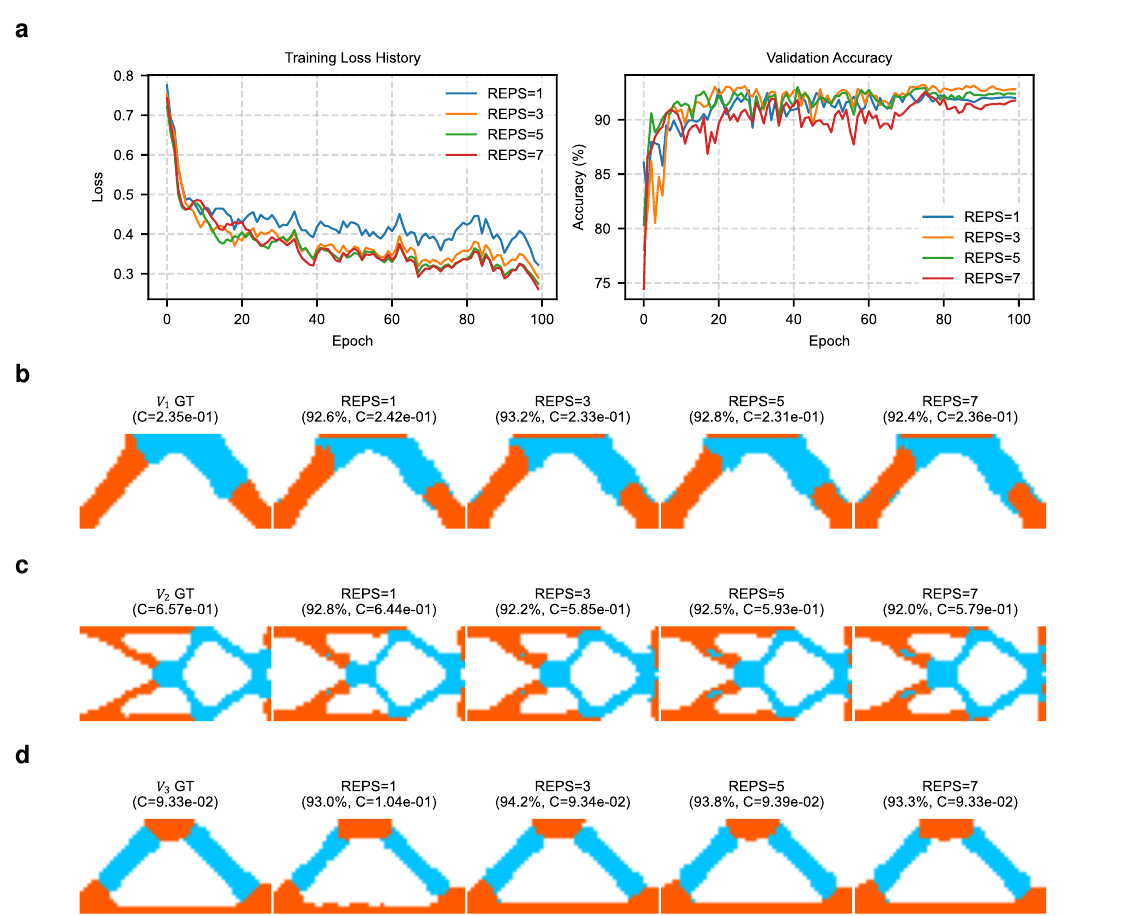}
\caption{\textbf{Effect of RealAmplitudes ansatz repetitions.} The ansatz repetition number was varied across $R\in\{1,3,5,7\}$. \textbf{a}, Training loss and mean validation accuracy over $V_1$--$V_3$. \textbf{b--d}, Validation predictions for $V_1$--$V_3$. REPS=3 gives a stable balance between validation performance and circuit depth.}
\label{fig:reps}
\addcontentsline{toc}{subsection}{Figure S3. Effect of RealAmplitudes ansatz repetitions}
\end{figure}

\begin{figure}[H]
\centering
\includegraphics[width=\linewidth, trim={0.0cm 12.0cm 0.0cm 0.0cm}, clip]{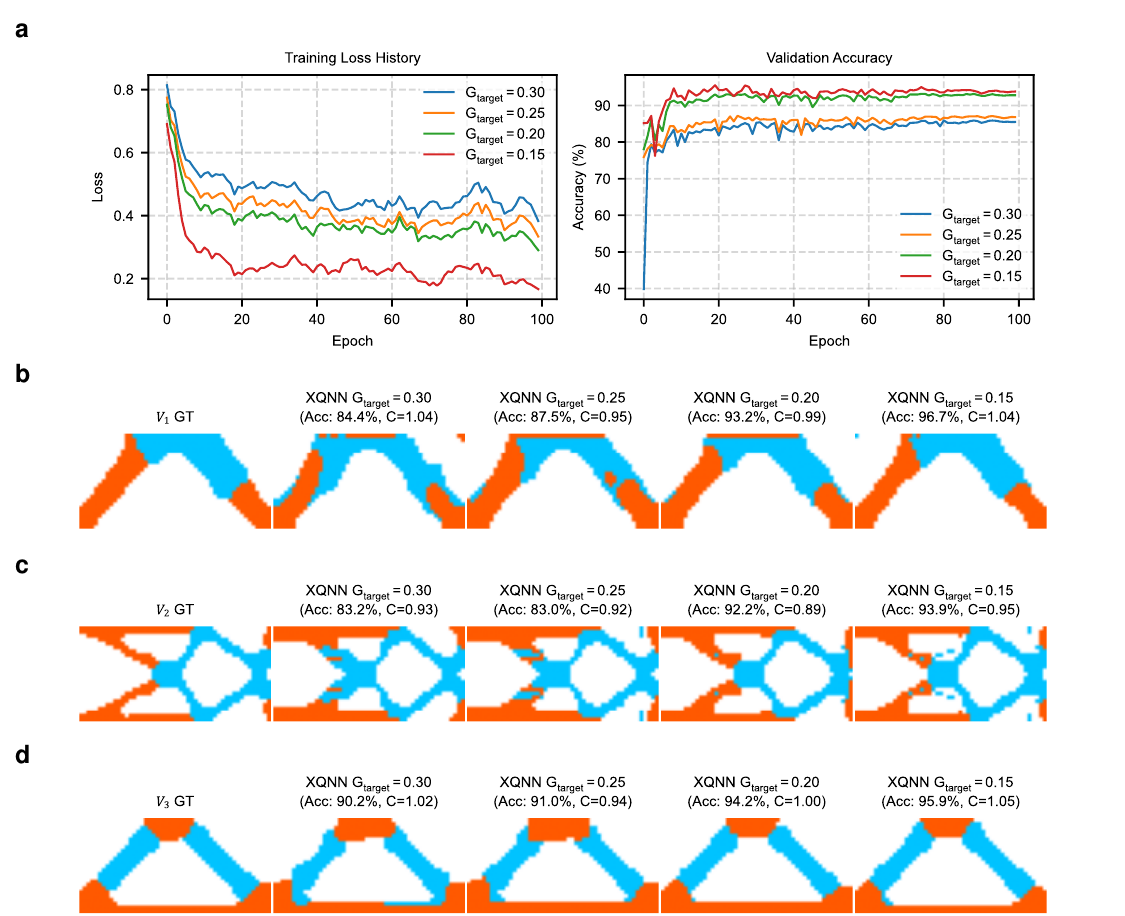}
\caption{\textbf{Effect of the prediction decision threshold $G_{\mathrm{target}}$.} Smaller values select later, sharper optimizer states and improve validation accuracy, whereas larger values stop earlier and preserve more acceleration. \textbf{a}, Training loss and mean validation accuracy over $V_1$--$V_3$. \textbf{b--d}, Validation predictions for $V_1$--$V_3$. The selected $G_{\mathrm{target}}$ values are chosen from this time--accuracy trade-off.}
\label{fig:gtarget}
\addcontentsline{toc}{subsection}{Figure S4. Effect of the prediction-decision threshold $G_{\rm target}$}
\end{figure}

The supplementary configuration studies support the default choices used in the main article using the validation problems $V_1$--$V_3$. The number of training designs is varied across $D\in\{1,3,5,7\}$ (Figure~\ref{fig:datasize}; Table~\ref{tab:s_datasize}): $D=1$ uses $T_1$ only, $D=3$ uses $T_1$--$T_3$ and is the default setting, $D=5$ additionally includes $O_1$ and $O_2$, and $D=7$ additionally includes $O_1$--$O_4$. The $D=3$ model already gives sufficient validation accuracy while preserving a strict separation between training and OOD testing. The $D=5$ and $D=7$ variants are used only to assess sensitivity to the number of training designs; the main reported OOD results use the default $D=3$ model, for which $O_1$--$O_4$ remain excluded from training and validation. Following the optimizer history motivation of Sukulthanasorn and Terada \cite{SukulthanasornTerada}, the optimizer history input size is varied across $N\in\{3,6,9,12\}$ physical descriptors before appending the Sobel boundary hint (Figure~\ref{fig:numqbit}; Table~\ref{tab:s_numqbit}). The validation accuracy over $V_1$--$V_3$ changes only modestly, and the $N=3$ setting can appear competitive when accuracy alone is considered. However, the loss decreases as more optimizer history descriptors are supplied, and the high-resolution transfer examples in Figure~\ref{fig:numqbit}e,f show that multiple milestones improve structural stability under altered two-material stiffness settings. In particular, the $N=3$ and $N=6$ inputs often miss the upper orange connector or produce thin, wavering blue members, whereas $N=9$ and $N=12$ better preserve connected material paths and hole shapes. The default therefore uses $N=9$, corresponding to $k_{\rm em}$, $k_{\rm em}+5$, and $k_{\rm pd}$, plus the Sobel boundary hint; $N=12$ adds $k_{\rm pd}-3$ but requires a wider 13-qubit statevector. The RealAmplitudes ansatz repetition number is varied across $R\in\{1,3,5,7\}$ (Figure~\ref{fig:reps}; Table~\ref{tab:s_reps}). The prediction decision threshold is varied across $G_{\rm target}\in\{0.25,0.20,0.15,0.10\}$ (Figure~\ref{fig:gtarget}; Table~\ref{tab:s_gtarget}). Validation accuracy averaged over $V_1$--$V_3$ is the primary model-selection metric, while the high-resolution examples are used to check whether the selected input history remains structurally meaningful beyond the validation mesh.

\phantomsection
\addcontentsline{toc}{subsection}{Supplementary Note 4. Observable-level analyses}
\subsection*{Supplementary Note 4. Observable-level analyses}

Margin quantum observable attribution maps (QOAM) are computed from the readout layer margin between the winning class $c_1$ and the nearest competing class $c_2$,
\begin{equation}
\Delta z_e=z_{e,c_1}-z_{e,c_2}
=(b_{c_1}-b_{c_2})+\sum_{j=0}^{9}Q_{e,j}\left(W_{j,c_1}-W_{j,c_2}\right).
\label{eq:s_margin}
\end{equation}
The signed channel contribution and non-negative QOAM magnitude are
\begin{equation}
m_{e,j}=Q_{e,j}\left(W_{j,c_1}-W_{j,c_2}\right),
\qquad
a_{e,j}=|m_{e,j}|,
\qquad
A_e=\sum_{j=0}^{9}a_{e,j}.
\label{eq:s_qoam}
\end{equation}
The readout layer is linear in $\mathbf{Q}_e$, so this attribution is exact for the final readout conditioned on the trained PQC output.

Feature-aligned channels are selected on validation datasets by correlating each quantum observable channel with strain energy, negative sensitivity, final material type label, and interface strength. For a target field $T_e$, the selected specialist channel is
\begin{equation}
j_T=\arg\max_j \left|\operatorname{corr}\left(Q_{e,j},T_e\right)\right| .
\label{eq:s_specialist}
\end{equation}
Here, interface strength is a local material-boundary proxy rather than a physical interface length. For a discrete final label field $y_e$, the two-dimensional interface strength target is computed by finite differences,
\begin{equation}
I_e=\sqrt{(\Delta_x y_e)^2+(\Delta_y y_e)^2},
\label{eq:s_interface}
\end{equation}
with the analogous finite-difference gradient used for voxelized three-dimensional fields. The specialist-channel analysis should be interpreted together with the readout and perturbation tests. Large single channel Pearson correlations indicate localized specialist observables, which occur for several energy, sensitivity, and material type targets in the simpler cases. Smaller correlations for the four-material and interface strength targets do not by themselves imply a loss of representation capacity. They indicate that the ZFeatureMap and entangling ansatz distribute the relevant information across several measured observables, so the final readout recovers the decision from collective observable responses rather than from one isolated qubit.

For erasure analysis, selected observable channels are replaced by neutral-reference values computed from a constant mid-range input vector. The erasure procedure is run on 400 shuffled elements, with 50 random-erasure trials. For sparsity analysis, channels are ranked globally by $\sum_c |W_{j,c}|$ and only the top $k$ channels are retained; all other channels are neutralized. Sparsity is run on 1000 shuffled elements, with random subsets evaluated over 50 trials. A plateau after retaining several top-ranked channels means that the dominant margin-carrying observables have already been included and that the remaining channels are mostly redundant or case-specific for the evaluated elements. For input feature ablation, the named descriptor group is removed from the input vector and the model is evaluated under the same training and validation protocol. The ablated groups are strain energy, sensitivity, density, and Sobel boundary hint.

\clearpage

\clearpage
\phantomsection
\addcontentsline{toc}{section}{Supplementary Figures}
\section*{Supplementary Figures}

\begin{figure}[H]
\centering
\includegraphics[width=\linewidth, trim={0.0cm 11.5cm 0.0cm 0.0cm}, clip]{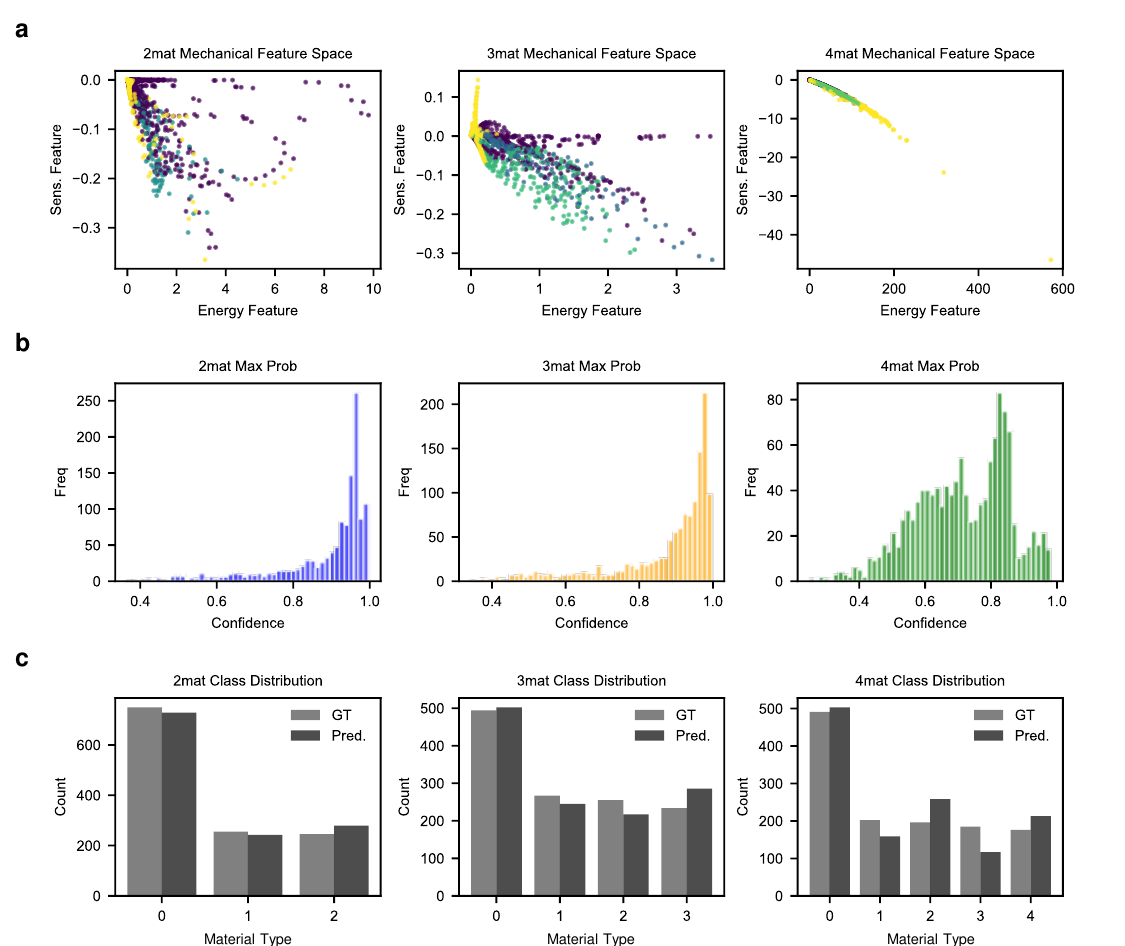}
\caption{\textbf{Mechanical feature space, classification confidence, and predicted material type distributions.} The diagnostic is evaluated for two-, three-, and four-material cases on an OOD test case. \textbf{a}, Strain energy versus sensitivity at the emergence stage. \textbf{b}, Maximum-softmax-probability histograms. \textbf{c}, GT and predicted material type frequencies.}
\label{fig:dist}
\addcontentsline{toc}{subsection}{Figure S5. Mechanical feature space, classification confidence, and predicted material type distributions}

\end{figure}

\begin{figure}[H]
\centering
\includegraphics[width=\linewidth, trim={0.0cm 7.5cm 0.0cm 0.0cm}, clip]{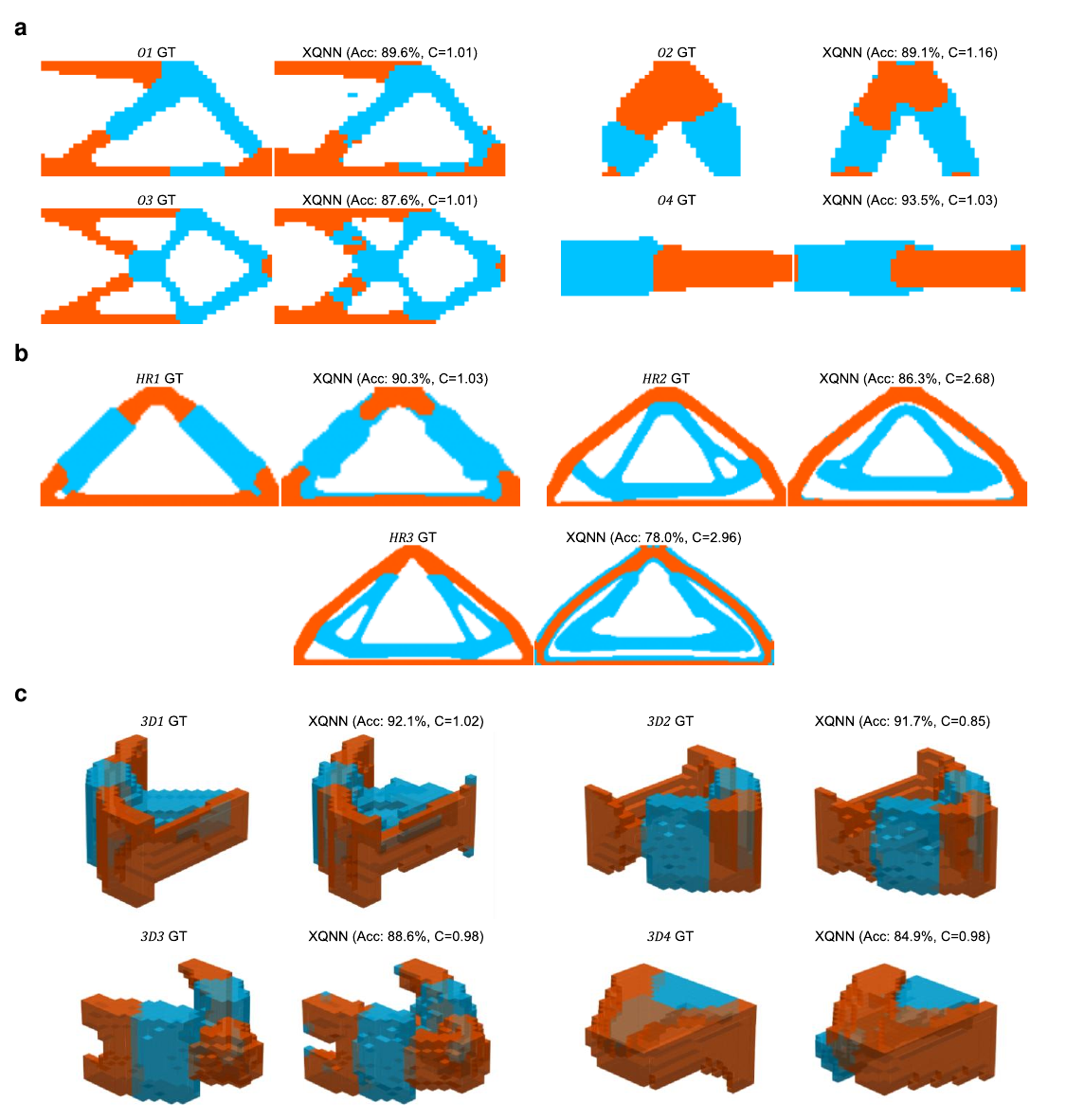}
\caption{\textbf{Full OOD, high-resolution, and three-dimensional predictions for the two-material setting.} \textbf{a}, GT and XQNN predictions for $O_1$--$O_4$. \textbf{b}, GT and XQNN predictions for $\mathit{HR}_1$--$\mathit{HR}_3$. \textbf{c}, GT and XQNN predictions for $\mathit{3D}_1$--$\mathit{3D}_4$. Reported $C$ values are normalized compliances $C_{\rm pred}/C_{\rm GT}$.}
\label{fig:full2mat}
\addcontentsline{toc}{subsection}{Figure S6. Full OOD, high-resolution, and three-dimensional predictions for the two-material setting}
\end{figure}

\begin{figure}[H]
\centering
\includegraphics[width=\linewidth, trim={0.0cm 7.5cm 0.0cm 0.0cm}, clip]{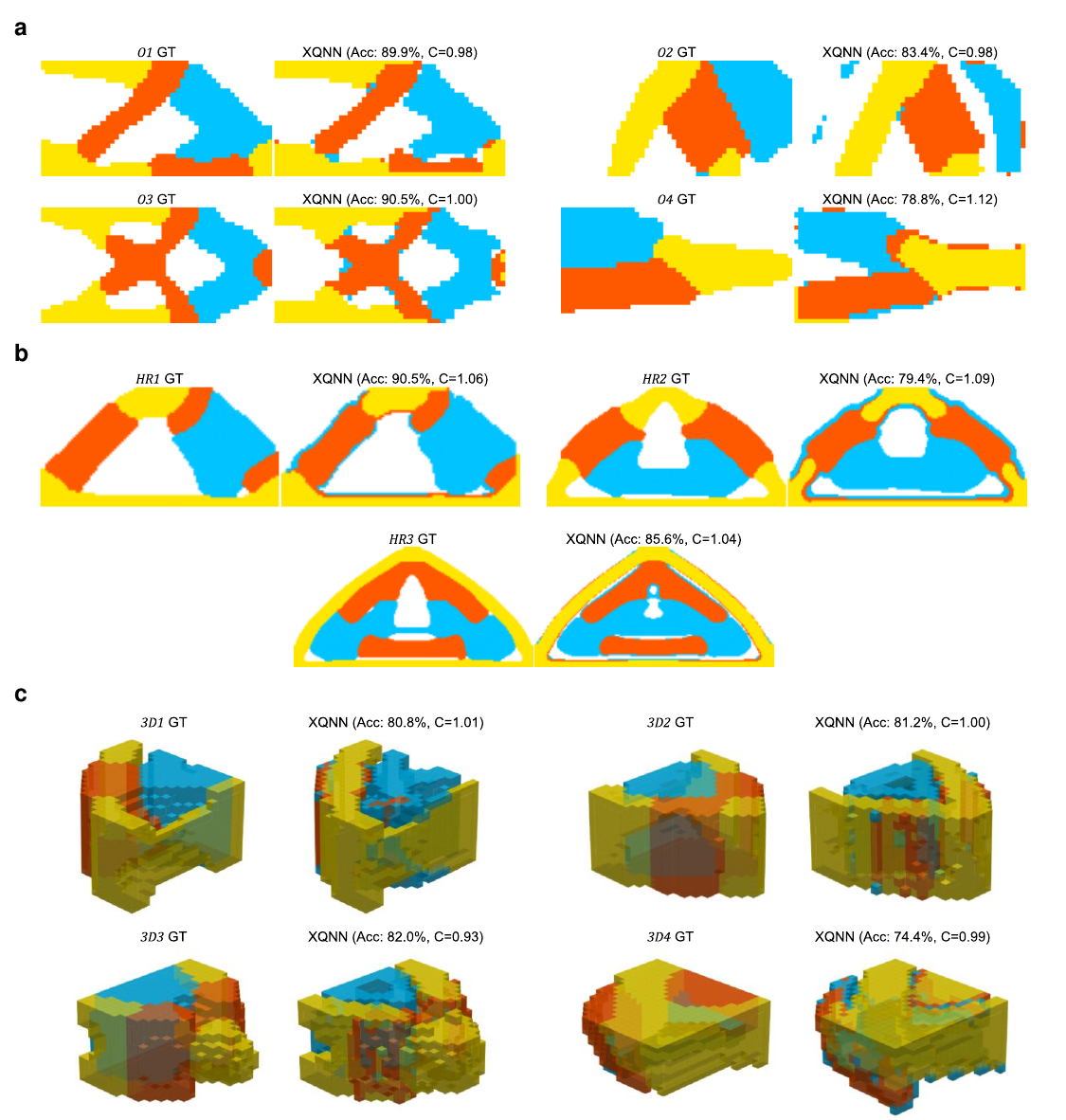}
\caption{\textbf{Full OOD, high-resolution, and three-dimensional predictions for the three-material setting.} \textbf{a}, GT and XQNN predictions for $O_1$--$O_4$. \textbf{b}, GT and XQNN predictions for $\mathit{HR}_1$--$\mathit{HR}_3$. \textbf{c}, GT and XQNN predictions for $\mathit{3D}_1$--$\mathit{3D}_4$. Reported $C$ values are normalized compliances $C_{\rm pred}/C_{\rm GT}$.}
\label{fig:full3mat}
\addcontentsline{toc}{subsection}{Figure S7. Full OOD, high-resolution, and three-dimensional predictions for the three-material setting}
\end{figure}

\begin{figure}[H]
\centering
\includegraphics[width=\linewidth, trim={0.0cm 7.5cm 0.0cm 0.0cm}, clip]{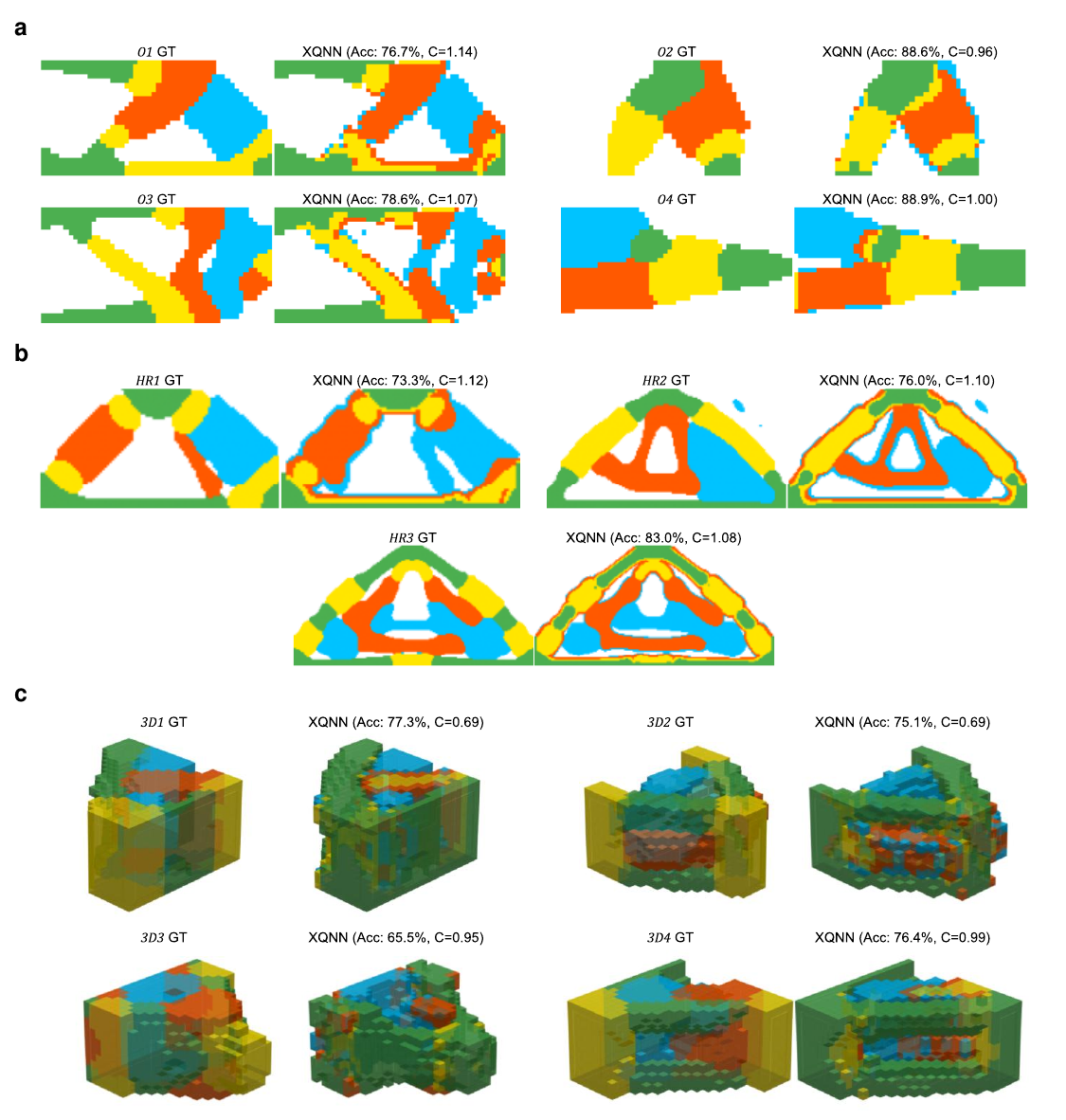}
\caption{\textbf{Full OOD, high-resolution, and three-dimensional predictions for the four-material setting.} \textbf{a}, GT and XQNN predictions for $O_1$--$O_4$. \textbf{b}, GT and XQNN predictions for $\mathit{HR}_1$--$\mathit{HR}_3$. \textbf{c}, GT and XQNN predictions for $\mathit{3D}_1$--$\mathit{3D}_4$. Reported $C$ values are normalized compliances $C_{\rm pred}/C_{\rm GT}$.}
\label{fig:full4mat}
\addcontentsline{toc}{subsection}{Figure S8. Full OOD, high-resolution, and three-dimensional predictions for the four-material setting}
\end{figure}

\clearpage

\phantomsection
\addcontentsline{toc}{section}{Supplementary Tables}
\section*{Supplementary Tables}

\begin{table}[htbp]
\footnotesize
\centering
\caption{\textbf{Dataset roles and boundary/loading definitions.} Training, validation, OOD, high-resolution, and three-dimensional transfer problems are disjoint. Training and validation diagrams are shown in main Figure~2; OOD, high-resolution, and three-dimensional transfer diagrams are shown in main Figure~3.}
\label{tab:s_dataset_roles}
\addcontentsline{toc}{subsection}{Table S1. Dataset roles}
\setlength{\tabcolsep}{3pt}
\renewcommand{\arraystretch}{1.18}
\begin{tabularx}{\linewidth}{@{}p{0.86in}p{0.82in}p{1.12in}Y@{}}
\toprule
Role & Problems & Used for & Boundary/loading description \\
\midrule
Training & $T_1$--$T_3$ & Parameter fitting & $50\times25$ cases: left-fixed right-edge vertical load; both-ends-fixed top-center vertical load; bottom-fixed upper-right horizontal load \\
Validation & $V_1$--$V_3$ & Model selection / early stopping & $50\times25$ cases: simply supported top-center vertical load; four-corner-fixed distributed top load; left-fixed two right-corner vertical loads \\
OOD test & $O_1$--$O_4$ & Final two-dimensional generalization test & Four additional $50\times25$ boundary/loading conditions defined in main Figure~3a and evaluated only after model selection \\
High-resolution & $\mathit{HR}_1$--$\mathit{HR}_3$ & Final mesh-transfer test & $100\times50$, $150\times75$, and $200\times100$ meshes under the high-resolution boundary/loading condition in main Figure~3b, excluded from training and validation \\
Three-dimensional transfer & $\mathit{3D}_1$--$\mathit{3D}_4$ & Final dimensional-transfer test & Four $20\times20\times10$ voxel boundary/loading cases defined in main Figure~3c, excluded from training and validation \\
\bottomrule
\end{tabularx}
\end{table}

\begin{table}[htbp]
\centering
\caption{\textbf{Ground-truth MMTO dataset-generation settings.} These settings correspond to the main two-dimensional datasets. The plotting colors follow the ordered stiffness list: void is white, and the first to fourth solid material types are shown as blue, orange, yellow, and green, respectively.}
\label{tab:s_dataset_settings}
\addcontentsline{toc}{subsection}{Table S2. Ground-truth MMTO dataset-generation settings}
\begin{tabular}{llll}
\toprule
Task & Solid Young's moduli & Volume constraints & Total solid volume \\
\midrule
2mat & $[1.0,2.0]$ & $V_1=V_2=0.20$ & 0.40 \\
3mat & $[1.0,2.0,5.0]$ & $V_m=0.20$ & 0.60 \\
4mat & $[1.0,2.0,5.0,10.0]$ & $V_m=0.15$ & 0.60 \\
\bottomrule
\end{tabular}
\end{table}

\begin{table}[htbp]
\centering
\caption{\textbf{Sensitivity to number of training designs.} Final training loss and mean validation accuracy over $V_1$--$V_3$ are reported for different training data compositions. The default setting is $D=3$.}
\label{tab:s_datasize}
\addcontentsline{toc}{subsection}{Table S3. Sensitivity to number of training designs}
\begin{tabularx}{\linewidth}{@{}cYcc@{}}
\toprule
Training designs $D$ & Training-data composition & Final loss & Validation accuracy \\
\midrule
1 & $T_1$ only & 0.3883 & 90.88\% \\
3 & $T_1$--$T_3$ (default) & 0.2957 & 92.80\% \\
5 & $T_1$--$T_3$ plus $O_1$ and $O_2$ & 0.5381 & 90.00\% \\
7 & $T_1$--$T_3$ plus $O_1$--$O_4$ & 0.4691 & 92.21\% \\
\bottomrule
\end{tabularx}
\end{table}

\begin{table}[htbp]
\centering
\caption{\textbf{Sensitivity to optimizer history input size and qubit count.} $N$ denotes the number of physical descriptors before the Sobel boundary hint is appended. The default setting is $N=9$, which uses the three MMA milestones $k_{\rm em}$, $k_{\rm em}+5$, and $k_{\rm pd}$.}
\label{tab:s_numqbit}
\addcontentsline{toc}{subsection}{Table S4. Sensitivity to optimizer history input size and qubit count}
\footnotesize
\setlength{\tabcolsep}{3pt}
\renewcommand{\arraystretch}{1.15}
\begin{tabularx}{\linewidth}{@{}ccYcc@{}}
\toprule
Physical descriptors $N$ & Total qubits & MMA milestones used & Final loss & Validation accuracy \\
\midrule
3 & 4 & $k_{\rm pd}$ only & 0.3727 & 93.68\% \\
6 & 7 & $k_{\rm em}$ and $k_{\rm pd}$ & 0.3280 & 92.11\% \\
9 & 10 & $k_{\rm em}$, $k_{\rm em}+5$, and $k_{\rm pd}$ (default) & 0.2957 & 92.80\% \\
12 & 13 & $k_{\rm em}$, $k_{\rm em}+5$, $k_{\rm pd}-3$, and $k_{\rm pd}$ & 0.2502 & 92.51\% \\
\bottomrule
\end{tabularx}
\end{table}

\begin{table}[htbp]
\centering
\caption{\textbf{Sensitivity to RealAmplitudes ansatz depth.} The default setting is REPS=3.}
\label{tab:s_reps}
\addcontentsline{toc}{subsection}{Table S5. Sensitivity to RealAmplitudes ansatz depth}
\begin{tabular}{cccc}
\toprule
REPS $R$ & Circuit parameters & Final loss & Validation accuracy \\
\midrule
1 & 20 & 0.3353 & 92.00\% \\
3 & 40 & 0.2957 & 92.80\% \\
5 & 60 & 0.2868 & 92.40\% \\
7 & 80 & 0.2689 & 91.76\% \\
\bottomrule
\end{tabular}
\end{table}

\begin{table}[htbp]
\centering
\caption{\textbf{Sensitivity to the prediction decision threshold $G_{\mathrm{target}}$.} The configuration study evaluates the time--accuracy trade-off associated with moving $k_{\rm pd}$ closer to the final optimized layout. Smaller thresholds use later optimizer states and typically improve validation accuracy, whereas larger thresholds preserve more early stopping acceleration.}
\label{tab:s_gtarget}
\addcontentsline{toc}{subsection}{Table S6. Sensitivity to the prediction-decision threshold $G_{\rm target}$}
\begin{tabular}{ccc}
\toprule
Case & Selected $G_{\mathrm{target}}$ & Rationale \\
\midrule
2mat & 0.20 & Retains acceleration while providing sufficient boundary clarity \\
3mat & 0.15 & Uses a later state to sharpen material type interfaces \\
4mat & 0.10 & Uses the sharpest selected state for the most interface-rich task \\
\bottomrule
\end{tabular}
\end{table}

\clearpage

\clearpage
\phantomsection
\addcontentsline{toc}{section}{Supplementary References}
\section*{Supplementary References}

\end{document}